\newif\ifisarxive
\isarxivetrue % use \isarxivetrue for ARXIV
\ifisarxive
    \documentclass{article}
    \pdfoutput=1
    \usepackage{arxiv}
\else
    \documentclass[12pt]{article}
    \usepackage[letterpaper, margin=1in]{geometry}
\fi

\usepackage[T1]{fontenc}
\usepackage{lmodern}
\usepackage[utf8]{inputenc}
\usepackage[english]{babel}	
\usepackage{graphicx, nicefrac}
\usepackage[super]{nth}

\usepackage{longtable}
\usepackage{siunitx}
\usepackage{pdflscape}
\usepackage{float}

\usepackage[pdftex, dvipsnames]{xcolor} 
\definecolor{mygold}{RGB}{205, 173, 0}
\definecolor{blue}{RGB}{0, 0, 213}
\definecolor{myblue}{RGB}{40, 75, 148}
\definecolor{gold3}{RGB}{205, 173, 0}

\usepackage{tikz}
\usetikzlibrary{shapes.geometric,arrows}

\usepackage{fancyhdr} 
\usepackage{hyperref}
\usepackage{amsmath}      
\usepackage{amsfonts} 
\usepackage{enumerate}
\usepackage{pdflscape} % landscape environment
\usepackage{mathtools}
\mathtoolsset{showonlyrefs,showmanualtags}

\usepackage{adjustbox}
\usepackage{multirow}
\usepackage{booktabs}
\usepackage{threeparttable}
\usepackage{makecell}
\usepackage{subcaption}

\newcolumntype{C}[1]{>{\centering\arraybackslash}m{#1}}
     
\usepackage{apacite}

\newcommand{\highlight}[1]{\textit{#1}}

\usepackage[acronym,section,nonumberlist]{glossaries}
\makenoidxglossaries
\newacronym{dif}{DIF}{Differential Item Functioning}
\newacronym{icc}{ICC}{Item Characteristic Curve}
\newacronym{iic}{IIC}{Item Information Curve}
\newacronym{lr}{LR}{Logistic regression}
\newacronym{irt}{IRT}{Item Response Theory}
\newacronym{mh}{MH}{Mantel-Haenszel}
\newacronym{pl}{PL}{Parameter Logistic}
\newacronym{sibtest}{SIBTEST}{Simultaneous Item Bias Test}
\newacronym{1pl}{1PL}{One-Parameter Logistic}
\newacronym{2pl}{2PL}{Two-Parameter Logistic}
\newacronym{3pl}{3PL}{Three-Parameter Logistic}
\newacronym{4pl}{4PL}{Four-Parameter Logistic}
\newacronym{ua}{UA}{Unsigned Area}
\newacronym{sa}{SA}{Signed Area}
\newacronym{cua}{CUA}{Closed Unsigned Area}
\newacronym{csa}{CSA}{Closed Signed Area}
\newacronym{ols}{OLS}{Ordinary Least Squares}
\newacronym{wsa}{WSA}{Weighted Signed Area}
\newacronym{wua}{WUA}{Weighted Unsigned Area}
\newacronym{wam}{WAM}{Weighted Area Measure}
\newacronym{ram}{RAM}{Rudner's Area Measure}
\newacronym{ets}{ETS}{Educational Testing Service}
\newacronym{hbsc}{HBSC}{Health Behaviour in School-aged Children}
\newacronym{msatb}{MSATB}{Medical School Admission Test in Biology}

\usepackage{appendix}

\DeclareRobustCommand\Rcode{\bgroup\@codex}
\def\@codex#1{\texorpdfstring%
{{\normalfont\ttfamily #1}}%
{#1}\egroup}

\providecommand{\keywords}[1]
{
  \small	
  \textbf{\textit{Keywords---}} #1
}

\usepackage{tocloft}
\newlength{\mylen}%1
\renewcommand{\cftfigpresnum}{\figurename\enspace}%2
\renewcommand{\cftfigaftersnum}{:}%3
\cftpagenumbersoff{figure}%remove list of figure page numbers
\usepackage{comment}

\title{Refining Effect-Size Measures and Classification for Differential Item Functioning: Toward Unified Guidelines Across Methods}

\ifisarxive
    \renewcommand{\shorttitle}{Toward Unified DIF Effect-Size Guidelines}
\fi
\author{Michaela Cichrová$^{1,2}$, Adéla Hladká$^{1}$, Patrícia Martinková$^{1,3}$\\
\small $^{1}$ Institute of Computer Science of the Czech Academy of Sciences, Prague, Czech Republic\\
\small $^{2}$ Faculty of Mathematics and Physics, Charles University, Prague, Czech Republic\\
\small $^{3}$ Faculty of Education, Charles University, Prague, Czech Republic\\
}
\date{June 10, 2026}

\begin{document}

%-------------------------------------------------------------------
% TITLE
%-------------------------------------------------------------------

\maketitle
%\thispagestyle{empty}

%-------------------------------------------------------------------
% ABSTRACT
%-------------------------------------------------------------------

\begin{abstract}

\noindent Differential Item Functioning (DIF) analysis is used to identify potentially biased items in multi-item measurements. In addition to testing the statistical significance, it is essential to evaluate the practical significance of DIF through effect-size measures. We review existing DIF effect-size measures and cut-off values used to classify the effect-size magnitudes for the Mantel-Haenszel test, SIBTEST, and model-based methods for binary items, and introduce a refinement of area-based effect-size measures. A simulation study is conducted to investigate the properties of these effect-size measures and existing classification guidelines, and to assess their comparative performance. The results indicate that some commonly used effect-size measures exhibit undesirable properties, including inconsistent classifications, systematic underestimation of the magnitude of the underlying DIF, and strong dependence on design factors. To address these issues, we introduce usage restrictions for some effect-size measures, revise cut-off values that unify results across different methods, and propose new cut-off values for area-based effect-size measures. The methods are demonstrated using two real data examples. Implementation is provided in the \texttt{R} software. 

\end{abstract}

%-------------------------------------------------------------------
% KEYWORDS
%-------------------------------------------------------------------

\noindent \keywords{Differential Item Functioning, Effect-Size Measures, Mantel-Haenszel Methods, \\ Logistic Regression}

\newpage

%-------------------------------------------------------------------
% MAIN TEXT
%-------------------------------------------------------------------

\clearpage
\pagenumbering{arabic} 

%-------------------------------------------------------------------
% INTRODUCTION
%-------------------------------------------------------------------
\section{INTRODUCTION}

Validation of newly developed multi-item measurement instruments, such as questionnaires or knowledge tests, involves several crucial steps. One key aspect is analyzing individual items and checking for \gls{dif}, which occurs when individuals from different groups or with different characteristics who have the same level of the underlying latent trait (e.g., ability or attitude) nevertheless have different probabilities of responding to an item in a particular way (e.g., endorsing a response category or answering correctly) \cite{dorans1993dif, osterlind2009dif, martinkova2023computational}. The presence of \gls{dif} indicates the existence of an underlying latent attribute that differs from the primary latent trait of interest. A close examination is required to determine whether such items should be reworded or removed from the test. If the secondary latent trait aligns with the construct of the test, it might be reasonable to keep the item; however, if it does not, removing such an item is essential to maintain the test's integrity and fairness, ensuring that the test measures the intended construct accurately for all groups \cite{martinkova2017checking}. 

Much of the methodological development in \gls{dif} detection has focused on settings with dichotomously scored items and comparisons between two groups, which remain the most common framework in applied research. Within this context, numerous methods have been developed over the past fifty years, including traditional approaches such as the \gls{mh} test \cite{holland1988differential} and \gls{sibtest} \cite{shealy1993model}, as well as model-based approaches such as logistic regression \cite{swaminathan1990detecting} and \gls{irt} models \cite{lord1968statistical}. However, different \gls{dif} detection methods can yield different conclusions about whether an item shows significant \gls{dif}, making the choice of method a critical but challenging decision. No single approach can be considered universally optimal; each has its own strengths and limitations, and is therefore more or less appropriate depending on the context \cite<see, e.g.,>[]{hidalgo2004differential}. In particular, model-based approaches provide more flexibility by allowing comparisons of nested models through submodel tests or information criteria, and they tend to be more powerful and robust than traditional methods, albeit at the cost of greater complexity \cite<e.g.,>[]{camilli1994methods, narayanon1996identification}. 

In addition to testing the statistical significance of \gls{dif}, it is also crucial to assess whether the detected \gls{dif} is of practical importance by quantifying its magnitude, given that the Type I error and power of statistical tests depend on the sample size \cite{suh2016effect, jodoin2001evaluating, roussos1996simulation, swaminathan1990detecting}. Thus, \gls{dif} analysis is often accompanied by a calculation of an appropriate measure of \highlight{\gls{dif} effect size} and its classification, usually as (A) "negligible", (B) "moderate", and (C) "large" \cite<see, e.g.,>[]{jodoin2001evaluating, potenza1995dif, zumbo1999handbook}. Such classification provides insight into the magnitude of \gls{dif} and helps to address the issue of Type I error inflation, which often arises in large datasets due to the increased statistical power \cite<see, e.g.,>{swaminathan1990detecting}, leading to a~detection of minimal differences. Such differences have almost no impact on total scores or affect just a tiny proportion of subjects and, therefore, should be classified as negligible \cite{wainer1993}. Items flagged for negligible \gls{dif} generally require no further inspection and can be kept in the test. Furthermore, because effect-size measures play a crucial role in evaluating practical significance, it is essential that their classification guidelines yield comparable results across different detection methods.

Historically, a variety of effect-size measures with various cut-off values for \gls{dif} detection methods have been proposed  \cite<see, e.g.,>[]{zumbo1997measure, jodoin2001evaluating, roussos1996simulation, zwick1989analysis, monahan2007odds}. Several classification guidelines were derived through simulation studies designed to align the cut-offs of one effect-size measure with those of another. However, the number of linked methods, scenarios, and replications considered was limited, possibly due to computational limitations at the time. In recent years, some effort has been made to update these cut-off values \cite<see, e.g.,>[]{weese2022reevaluating}, yet these efforts have not fully addressed the need for a more thorough review, as they often rely on values established in earlier, less comprehensive simulations. 

To address this gap, this article revisits the classification guidelines for multiple \gls{dif} effect-size measures in a complex simulation study, with the aim of improving consistency and interpretability across methods. In particular, we argue that \gls{dif} classification should not rely on outdated cut-off values. These were often derived from relatively small-scale simulation studies \cite<e.g.,>[]{jodoin2001evaluating} or have not been systematically validated at the time of their proposal \cite<e.g.,>[]{monahan2007odds, zumbo1997measure}. As a result, they may not generalize well to broader conditions. Instead, we promote a systematic mapping of effect-size measures across different \gls{dif} detection approaches, enabling a more coherent interpretation of results obtained from traditional and model-based methods. To facilitate this unification, we introduce a standardized version of area-based \gls{dif} effect sizes that can be directly linked to traditional, established metrics. In addition, we discuss the limitations and problematic aspects of existing effect-size measures and propose specific usage restrictions to address them. The investigation is conducted within the commonly studied framework of dichotomous items and a binary group membership variable, and the proposed unified cut-off values are intended for this setting. 

The paper proceeds as follows. In Section~\ref{sec:methods}, we provide an overview of traditional and model-based \gls{dif} detection methods, corresponding effect-size measures, existing classification guidelines, and proposed standardized versions of the area-based measures. In Section~\ref{sec:implementation}, we describe the implementation of the methods. In Section~\ref{sec:simulation}, we present the design and results of the simulation study, which serve as the basis for the newly proposed \gls{dif} effect-size classification guidelines. In Section~\ref{sec:realdata}, we illustrate the methods on two real data examples.
Section~\ref{sec:discussion} provides a discussion.

%-------------------------------------------------------------------
% METHODS
%----------------------------------------------------------------
\section{METHODS}\label{sec:methods}

%-------------------------------------------------------------------
\subsection{DIF detection methods and effect-size measures}

When examining \gls{dif} with respect to a single binary grouping variable, it is commonly distinguished as either \highlight{uniform} or \highlight{non-uniform}. In the case of uniform \gls{dif}, one group is consistently favored across all ability levels, while in non-uniform \gls{dif}, the advantaged group varies depending on the ability level \cite{hambleton1993advances}. Conventionally, the reference group typically refers to the standard group presumed to have an advantage, while the focal group is analyzed to identify potential disadvantages in comparison.

To detect these two forms of \gls{dif}, a range of approaches can be used. Among the most widely applied are traditional techniques, such as the \gls{mh} test, which is primarily designed for identifying uniform \gls{dif}, and the \gls{sibtest} method, along with its crossing \gls{sibtest} extension, which can detect both uniform and non-uniform \gls{dif}. Model-based approaches, such as \gls{lr} and \gls{irt} models, offer greater flexibility and can also accommodate both types of \gls{dif}. Given their widespread use in an applied measurement context, we focus on these well-established procedures in the present study.

%-----------------------
\subsubsection{Mantel-Haenszel test}

A traditional method for detecting uniform \gls{dif} is the \gls{mh} test \cite{holland1988differential}, which was originally introduced by \citeA{mantel1959statistical} in the field of biostatistics. The \gls{mh} test can be seen as an extension of the $\chi^2$ test of independence in the contingency table by considering distinct contingency tables for each value of the total score $k = 0, \dots, K$, containing counts of respondents with a~total score $k$ that answered correctly ($n_{i01k}$ for reference group, $n_{i11k}$ for focal group), and incorrectly ($n_{i00k}$ for reference group, $n_{i10k}$ for focal group). It is also possible to use other matching criteria, besides the total score. The corresponding \gls{mh} test statistic, with $\chi^2_1$ distribution under the null hypothesis of no \gls{dif}, is given by 
\begin{equation*}
    \text{MH}_i = \frac{\left(\Big\vert \sum_{k = 0}^{K}  \left(n_{i01k} - \frac{n_{i0+k} n_{i+1k}}{ n_{i++k}}\right) \Big\vert - 0.5 \right)^{2} }{\sum_{k = 0}^{K} \frac{n_{i0+k} n_{i1+k} n_{i+0k} n_{i+1k}}{n_{i++k}^2 \left( n_{i++k} - 1 \right)}},
\end{equation*}
where $n_{i+1k} = n_{i01k} + n_{i11k}$ is the total number of correct answers to item $i$ in both groups. Similarly, $n_{i+0k} = n_{i00k} + n_{i10k}$ denotes the total number of incorrect answers in both groups. Next, $n_{i0+k} = n_{i00k} + n_{i01k}$ and $n_{i1+k} = n_{i10k} + n_{i11k}$ are the numbers of respondents in the reference and focal group, respectively. Finally, $n_{i++k} = n_{i00k} + n_{i01k} + n_{i10k} + n_{i11k}$ is the total number of respondents.  

The effect magnitude can be quantified using the weighted average of the odds ratios across all levels of the total score, as proposed by \citeA{holland1985alternate}:
\begin{equation*}\label{eq:mh:alpha}
    \alpha_{\text{MH}i} = \frac{\sum_{k = 0}^{K} \frac{n_{i01k} n_{i10k}}{n_{i++k}}}{\sum_{k=0}^{K} \frac{n_{i00k} n_{i11k}}{n_{i++k}}}.
\end{equation*}
Values of $\alpha_{\text{MH}i}$ close to 1 indicate no evidence of \gls{dif}. Values lower than 1 suggest a bias in favor of the focal group. In contrast, values greater than 1 indicate a bias in favor of the reference group. 

To simplify interpretation, $\alpha_{\text{MH}i}$ is often log-transformed and rescaled as per \gls{ets} convention into
\begin{equation}\label{eq:mh:delta}
    \Delta_{\text{MH}_i} = - 2.35\, \log \left( \alpha_{\text{MH}i}\right),
\end{equation}
see \citeA{holland1985alternate}. In this transformation, values of $\Delta_{\text{MH}i}$ close to 0 imply the absence of \gls{dif}. Positive values of  $\Delta_{\text{MH}i}$ $(\Delta_{\text{MH}i} > 0)$ indicate that the focal group is favored, while negative values $(\Delta_{\text{MH}i} < 0)$ suggest a bias favoring the reference group. 

The commonly used cut-off values to classify the effect-size measure $\vert \Delta_{\text{MH}i} \vert$ as negligible, moderate, and large are 1 and 1.5, respectively, as follows: 
\begin{align}\label{eq:mh:es}
    \vert \Delta_{\text{MH}i} \vert \in 
    \begin{cases}
        [0, 1) & \text{negligible (category A)},\\
        [1, 1.5) & \text{moderate (category B)},\\
        [1.5, \infty) & \text{large (category C)},\\
    \end{cases}
\end{align}
for more details, see \citeA{zwick1989analysis}.

The interpretation of the $\Delta_{\text{MH}}$ metric and its classification thresholds can be expressed in terms of odds ratios. A value of $\Delta_{\text{MH}} = -1$, corresponding to the threshold for moderate \gls{dif}, implies $\alpha_{\text{MH}} = \exp(1 / 2.35) \approx 1.53$. This means that the reference group is about 1.53 times as likely to respond correctly (in terms of odds) as the reference group. In other words, if the reference group has a 50\% probability of answering this item correctly at a given ability level, the focal group would have a probability of about 39\%. A value of $\Delta_{\text{MH}} = 1$ indicates the same magnitude of \gls{dif} but in the opposite direction, with the focal group being advantaged. 

Similarly, a value of $\Delta_{\text{MH}} = -1.5$, corresponding to the threshold for large \gls{dif}, yields $\alpha_{\text{MH}} = \exp(1.5 / 2.35) \approx 1.89$. In this case, the reference group is approximately 1.89 times as likely to answer correctly as the focal group. Under the same reference-group probability of 50\% at a given ability level, the focal-group probability would increase to approximately 35\%. Again, $\Delta_{\text{MH}} = 1.5$ represents the same magnitude of \gls{dif} in the opposite direction.

%-----------------------
\subsubsection{SIBTEST}

Another classical method for detecting uniform \gls{dif} is \gls{sibtest} \cite{shealy1993model}. This approach evaluates differences in the expected probabilities of correct responses between the focal and reference groups, assuming equivalent total test score distributions. The test statistic for item $i$, $\hat{\beta}_i,$ is constructed by stratifying respondents based on their total scores and computing the weighted average of the differences in the mean score of the items between the focal and reference groups within each stratum. These strata are typically defined using the composite score of all item scores except the one being analyzed, denoted as: 
\begin{equation*}
    X^{\left(-i\right)} = \sum_{j \neq i, j = 1}^{K} Y_j,
\end{equation*}
where $Y_j \in \lbrace 0, 1 \rbrace$ represents the score for item $j$, and consequently $X^{\left(-i\right)} \in \lbrace 0,\dots, K - 1 \rbrace $. The weights used in the estimation are the proportions $\pi_k$ of respondents with composite score $k,$ such that $X^{\left(-i\right)} = k.$ Specifically,
\begin{equation}\label{eq:sibtest:hatbeta}
    \hat{\beta}_i = \sum_{k = 0}^{K-1} \pi_k\left(  \overline{Y}_{i0k} - \overline{Y}_{i1k} \right),
\end{equation} 
where $\overline{Y}_{i0k}$ and $\overline{Y}_{i1k}$ are the mean scores for item $i$ in the reference and focal groups, respectively, within each stratum.

Under the assumption that the target ability distribution is the same for both the focal and the reference groups, \citeA{shealy1993model} provided a formula for $\hat{\sigma} ( \hat{\beta}_i )$, the estimated standard deviation of $\hat{\beta}_i$. They also proposed using a test statistic 
\begin{equation*}
    B_i = \frac{ \hat{\beta}_i }{ \hat{\sigma} ( \hat{\beta}_i )}
\end{equation*} 
to test for the presence of \gls{dif}. However, as the assumption of identical ability distributions is often unrealistic, \citeA{shealy1993model} further proposed replacing the mean item scores $\overline{Y}_{i0k}$ and $\overline{Y}_{i1k}$ by their regression-based estimates. 

A modified version of the \gls{sibtest} method, called the crossing \gls{sibtest}, was introduced by \citeA{li1996new} to test for non-uniform \gls{dif}. The idea is to modify the test statistic in as \begin{equation}\label{eq:sibtest:hatbeta:nunif}
    \hat{\beta}_i = \sum_{k = 0}^{k_c - 1} \hat{\pi}_k ( \bar{Y}_{i0k} -\bar{Y}_{i1k}) - \sum_{k = k_c}^{K} \hat{\pi}_k ( \bar{Y}_{i0k} -\bar{Y}_{i1k}),
\end{equation}
where $k_c$ is an estimate of the crossing point, estimated in advance. 

\citeA{roussos1996simulation} introduced the $\vert \hat{\beta}_i \vert$ statistic to classify the magnitude of detected uniform \gls{dif}. Values of $\vert \hat{\beta}_i \vert$ close to 0 suggest the absence of \gls{dif}, with cut-off values of 0.059 and 0.088 used to distinguish between negligible, moderate, and large \gls{dif}. Specifically, 
\begin{align}\label{eq:sibtest:hatbeta:thresholds}
    \vert \hat{\beta}_i \vert \in 
    \begin{cases}
        [0, 0.059) & \text{negligible (category A)},\\
        [0.059, 0.088) & \text{moderate (category B)},\\
        [0.088, \infty) & \text{large (category C)}.
    \end{cases}
\end{align}
These thresholds were derived from the cut-offs for the $\vert \Delta_{\text{MH}i} \vert$ statistic in  \eqref{eq:mh:es}. The derivation was based on a strong correlation between  $|\Delta_{\text{MH}i}|$ and $|\hat{\beta}_i|$, as demonstrated through a small-scale simulation study for uniform items. However, the cut-off values can also be used for non-uniform \gls{dif} \cite{chalmers2018improving}.

\citeA{weese2022reevaluating} updated the cut-off values for specific underlying \gls{irt} models, deriving them from the thresholds established for the $\vert \Delta_{\text{MH}i} \vert$ statistic in \eqref{eq:mh:es}. For the \gls{2pl} \gls{irt} model \cite{birnbaum1968statistical}, the revised thresholds were 0.062 and 0.09, while for the \gls{3pl} \gls{irt} model, the updated values were 0.069 and 0.102. 

%-----------------------
\subsubsection{Logistic regression and its extensions}\label{sec:methods:lr}

Model-based approaches involve building a group-specific regression model that incorporates \gls{dif} parameters, which allow for estimation of how the item functioning differs between the focal and reference groups. This model is then compared to a submodel where the \gls{dif} parameters are constrained to be equal across groups, allowing for a test of whether \gls{dif} is present.

\citeA{swaminathan1990detecting} proposed using a group-specific \gls{lr} model to detect \gls{dif}. The probability of a correct answer on item $i$ by a respondent $p$ is modeled using the following covariates: the group membership ($G_p$; typically $0 =$ reference group, $1 =$ focal group), the ability matching criterion ($\theta_p$) (observed ability estimate, such as the standardized total score), and their interaction. The formula for the log odds of answering item $i$ correctly is given by: 
\begin{align}\label{eq:lr}
    \log \left( \frac{\mathsf{P} \left( Y_{pi} = 1 \vert \theta_p,\, G_p \right)}{1 - \mathsf{P} \left( Y_{pi} = 1 \vert \theta_p,\, G_p \right)}\right) = \beta_{i0} + \beta_{i1} \theta_{p} + \beta_{i2} G_{p} + \beta_{i3} \theta_{p} G_{p}.
\end{align}
To test for \gls{dif} in item $i$, a series of model-submodel tests can be used. The \gls{lr} approach can determine whether the \gls{dif} is uniform or non-uniform (see Table \ref{tab:tests} for details). The coefficients are estimated via maximum likelihood using the iteratively re-weighted least squares algorithm, and hypothesis testing is typically performed using likelihood ratio tests.

\begin{table}[H]
    \centering
     \caption{Submodel tests to detect different types of DIF.}
    \begin{tabular}{l l l}
        \toprule
        \multirow{2}{*}{DIF Type} & \multicolumn{2}{c}{Hypothesis} \\ \cmidrule(lr){2-3}
         & \multicolumn{1}{c}{Null $H_0$} & \multicolumn{1}{c}{Alternative $H_1$} \\ 
        \midrule
        Any  & $\beta_{i2} = 0 \text{ and } \beta_{i3} = 0$ & $\beta_{i2} \neq 0 \text{ or } \beta_{i3} \neq 0$ \\
        Uniform & $\beta_{i2} = 0 \mid \beta_{i3} = 0$ & $\beta_{i2} \neq 0 \mid \beta_{i3} = 0$ \\
        Non-uniform & $\beta_{i3} = 0$ & $\beta_{i3} \neq 0$ \\
        \bottomrule
    \end{tabular}
    \label{tab:tests}
\end{table}

The equation \eqref{eq:lr} can be rewritten using an \gls{irt} parametrization
\begin{align}\label{eq:lrirt}
    \log \left( \frac{\mathsf{P} \left( Y_{pi} = 1 \vert \theta_p,\, G_p \right)}{1 - \mathsf{P} \left( Y_{pi} = 1 \vert \theta_p,\, G_p \right)}\right) = (a_i + a_{\text{DIF}_i} G_p)(\theta_p - b_i - b_{\text{DIF}_i} G_p),
\end{align}
where $b_i$ and $b_i + b_{\text{DIF}_i}$ are the difficulty parameters of the reference and focal group, and $a_i$ and $a_i + a_{\text{DIF}_i}$ are the discrimination parameters of the reference and focal group for the item $i$, respectively.

\vspace{2ex}

Over the years, numerous effect-size measures have been developed and used to quantify identified \gls{dif}. Below is a list of commonly used measures.  

\paragraph{$\mathbf{ \Delta R^2}$: Difference of Nagelkerke's (pseudo) $\mathbf{R^2}$.} 
The standardized version of Nagelkerke's $R^2$ is defined as 
\begin{align*}
    R^2_{i, \text{std}} = \frac{R^2_{i}}{R^2_{i, \text{max}}},
\end{align*}
where
\begin{align*}
    R^2_i = 1 - \exp{\left( \frac{\text{dev}_i -\text{dev}_{i, \text{null}}}{n}\right)}
\end{align*}
is a non-standardized version of Nagelkerke's $R^2$ with $\text{dev}_i$ standing for the deviance of the fitted model and $\text{dev}_{i, \text{null}}$ for the deviance of the null model, i.e., a model with just an intercept. Further,  
\begin{align*}
    R^2_{i, \text{max}} = 1 - \exp{\left( \frac{-\text{dev}_{i, \text{null}}}{n}\right)}
\end{align*}
is the $R^2$ of the null model. \citeA{zumbo1997measure} proposed measuring effect size as the difference in Nagelkerke's $R^2$ between a model and its corresponding submodel, denoted as $\Delta R^2_{i, \text{std}}$ (see Table \ref{tab:tests}).  

Historically, \citeA{zumbo1997measure} suggested cut-off values of 0.13 and 0.26 using the bounds introduced by~\citeA{cohen1992power} for multivariate correlation measures. Specifically,
\begin{align}
    \Delta R^2_{i, \text{std}} = R^2_{i, \text{std}, \text{model}} - R^2_{i, \text{std}, \text{submodel}} \in 
    \begin{cases}
        [0, 0.13) & \text{negligible (category A)},\\
        [0.13, 0.26) & \text{moderate (category B)},\\
        [0.26, \infty) & \text{large (category C)}. 
    \end{cases}
\end{align}
\citeA{jodoin2001evaluating} derived improved thresholds of 0.035 and 0.07 through a simulation study using cut-off values proposed for \gls{sibtest} by \citeA{roussos1996simulation} by using cubic regression. Explicitly, 
\begin{align}\label{eq:lr:R2diff}
    \Delta R^2_{i, \text{std}} \in 
    \begin{cases}
        [0, 0.035) & \text{negligible (category A)},\\
        [0.035, 0.07) & \text{moderate (category B)},\\
        [0.07, \infty) & \text{large (category C)}. 
    \end{cases}
\end{align}

\paragraph{$\mathbf{\Delta_{\text{LR}}}$: Transformed reference-to-focal odds ratio.} \citeA{monahan2007odds} offered several measures to quantify the magnitude of uniform \gls{dif}. One such measure was the reference-to-focal conditional odds ratio: \begin{align}\label{eq:lr:alphaLR}
    \hat{\alpha}_{\text{LR}_i} = \exp (-\hat{\beta}_{i2}),
\end{align}
where $\hat{\beta}_{i2}$ is the estimated effect of group membership in the model \eqref{eq:lr} without its interaction with the matching criterion (i.e., $\beta_{i3} = 0$). Odds ratios are the most natural effect sizes for \gls{lr} models \cite<>[Chapter 5]{agresti2002categorical}. To facilitate comparison with the \gls{mh} effect-size metric \eqref{eq:mh:delta}, \citeA{monahan2007odds} further suggested transforming $\hat{\alpha}_{\text{LR}_i}$ as:
\begin{align}\label{eq:lr:deltaLR}
    \hat{\Delta}_{\text{LR}_i} = -2.35\, \text{log} \left(\hat{\alpha}_{\text{LR}_i} \right) =  2.35\, \hat{\beta}_{i2},
\end{align} 
using the same cut-off values as for the \gls{mh} case specified in \eqref{eq:mh:es}. Under the \gls{irt} parametrization \eqref{eq:lrirt}, $\hat{\Delta}_{\text{LR}_i}$ corresponds to $-2.35\, a_i\, b_{i\text{DIF}}$, indicating that the resulting effect size depends not only on shift in item difficulty, but also on the item's discrimination power. 

\paragraph{LR-P-DIF: Logistic regression p-differences.} \citeA{monahan2007odds} also proposed converting $\hat{\alpha}_{\text{LR}_i}$ into a \textit{p} metric for classification of uniform \gls{dif}: 
\begin{align}\label{eq:lr:LRPDIF}
    \text{LR-P-DIF}_i = \pi_{i1} - \pi_{i0}^\Psi,
\end{align} 
where $\pi_{i1}$ is the proportion of correct answers for the focal group and $\pi_{i0}^\Psi = \frac{\hat{\alpha}_{\text{LR}_i}\pi_{i1}}{(1-\pi_{i1}) + \hat{\alpha}_{\text{LR}_i} \pi_{i1}}$ is the predicted proportion of correct answers for focal group based on $\hat{\alpha}_{\text{LR}_i}$. The same formula was applied for the \gls{mh} test by \citeA{dorans1991constructed}. \citeA{monahan2007odds} recommended using historically established cut-off values for \textit{p} metrics of 0.05 and 0.1 for $\vert \text{LR-P-DIF}_i \vert$ to distinguish negligible, moderate, and large DIF effect sizes. Specifically, 
\begin{align}\label{eq:lr:pdif}
\vert \text{LR-P-DIF}_i \vert  \in 
    \begin{cases}
        [0, 0.05) & \text{negligible (category A)},\\
        [0.05, 0.1) & \text{moderate (category B)},\\
        [0.1, \infty) & \text{large (category C)}.
    \end{cases}
\end{align}

\paragraph{LR-STD-P-DIF: Logistic regression standardized p-differences.} Another option to classify \gls{dif} magnitude is a standardized \textit{p} metric LR-STD-P-DIF \cite{dorans1986demonstrating, dorans1993dif}, which for the \gls{lr} takes the following form: 
\begin{align}\label{eq:lr:LRSTDPDIF}
    \text{LR-STD-P-DIF}_i = \frac{\sum_{\theta = 0}^{K} w_{\theta} \left( \pi_{i1\theta} - \pi_{i0\theta} \right)}{\sum_{\theta = 0}^K w_\theta}, 
\end{align} 
where the proportions of correct answers for the focal and reference groups $\pi_{i1\theta}$ and $\pi_{i0\theta}$ with total score $\theta$ are predicted from the model \eqref{eq:lr}. There are several options for choosing the weights \( w_\theta \)
. \textbf{Total weights} correspond to the number of respondents in each stratum (\(N_\theta\)) \cite{monahan2007odds, dorans1993dif}. \textbf{Reference weights} use the number of reference group respondents in each stratum (\(N_{1\theta}\)), while \textbf{focal weights} are based on the number of focal group respondents (\(N_{0\theta}\)) \cite{dorans1986demonstrating}. \textbf{Cochran's weights} \cite{dorans1993dif} are defined as 
\begin{align*}
    w_\theta = \frac{N_{1\theta}N_{0\theta}}{N_{\theta}}.
\end{align*} 
Lastly, \textbf{uniform weights} set \( w_\theta = 1 \) for all $\theta = 0, \dots, K$ \cite{dorans1993dif}.

As for the $|\text{LR-P-DIF}_i|$ measure, \citeA{monahan2007odds} recommended using thresholds of 0.05 and 0.1 for $\vert \text{LR-STD-P-DIF}_i \vert$ to distinguish negligible, moderate, and large \gls{dif} effect sizes, analogous to \eqref{eq:lr:pdif}. 

%-----------------------
\subsubsection{IRT-based methods}\label{generalized_lr}

\gls{dif} can be examined using group-specific \gls{irt} models analogous to the \gls{lr} method presented in Equation \eqref{eq:lrirt}. The main difference between these two approaches is that in the \gls{irt} framework, both item parameters and the latent ability $\theta$ are estimated simultaneously, rather than conditioning on a pre-estimated ability, e.g., the total score. 

The \gls{2pl} model for both \gls{lr} and \gls{irt} frameworks can be extended to incorporate additional response behavior such as guessing (lower asymptote, parameter $c$) and inattention (upper asymptote, parameter $d$) \cite{birnbaum1968statistical, barton1981upper, drabinova2017detection, hladka2020difnlr}. In the \gls{irt} setting, this leads to the \gls{4pl} \gls{irt} model \cite{barton1981upper}:
\begin{equation}
\begin{aligned}\label{eq:irt:4PL}
    \mathsf{P}(Y_{pi} = 1 \vert \theta, G_p) = c_i + c_{\text{DIF}_i} G_p + (d_i &+ d_{\text{DIF}_i}G_p - c_i - c_{\text{DIF}_i} G_p )\, \\
    &\cdot \frac{\exp((a_i + a_{\text{DIF}_i}G_p)(\theta - b_i - b_{\text{DIF}_i}G_p))}{1 + \exp((a_i + a_{\text{DIF}_i}G_p)(\theta - b_i - b_{\text{DIF}_i)G_p)}}. 
\end{aligned} 
\end{equation}
By setting $d_i = 1$ and $d_{\text{DIF}_i} = 0$, this model reduces to the \gls{3pl} \gls{irt} model \cite{birnbaum1968statistical}.

\vspace{1em}

Several \gls{dif} effect-size measures have been introduced in the context of the \gls{irt} model, which may also be used for the \gls{lr} model \eqref{eq:lrirt}. 

\paragraph{SA, UA: Signed and unsigned area measures.}
A natural effect-size measure for \gls{dif} is the area between the \glspl{icc} of the focal and reference groups, $F_{i1} \left( \theta \right) = \mathsf{P}(Y_i = 1 \vert \theta, G = 1)$ and $F_{i0}  \left( \theta \right) = \mathsf{P}(Y_i = 1 \vert \theta, G = 0)$, as originally suggested by \citeA{rudner1977approach}. We consider two related measures: the \gls{sa} and the \gls{ua}, defined as:
\begin{align}
    \text{SA}_i \label{eq:irt:es:SA} =&\ \int_{-\infty}^{\infty} \left(  F_{i0}  \left( \theta \right) - F_{i1}  \left( \theta \right)\right) \, \mathrm{d}\theta,\\
    \text{UA}_i = \label{eq:irt:es:UA} &\ \int_{-\infty}^{\infty} \vert F_{i0}  \left( \theta \right) - F_{i1}  \left( \theta \right) \vert \, \mathrm{d}\theta. 
\end{align}

\citeA{rudner1977approach} proposed estimating the area between \glspl{icc} using successive rectangles. In this approach, the total area is approximated by summing the contribution of rectangles of width 0.005 units over the $\theta$ range from $-5$ to $5$. 

Later, \citeA{raju1988area} derived explicit formulae for the signed and unsigned area in terms of \gls{irt} model parameters. For instance, under the \gls{2pl} model~\eqref{eq:lrirt} within the corresponding \gls{irt} framework, the formulae are
\begin{align}
    \text{SA}_i \label{SA}&= b_i + b_{i\text{DIF}} - b_i = b_{i\text{DIF}}, \\
    \text{UA}_i \label{UA}&= 
    \begin{cases} \,
        \vert b_{i\text{DIF}} \vert & \text{if } a_{i\text{DIF}} = 0 ,\\[10pt]
        \left| \frac{2 a_{i\text{DIF}} }{ a_i \left( a_i + a_{i\text{DIF}}\right) } 
        \ln \left( 1 + \exp \left( \frac{ b_{i\text{DIF}} a_i \left( a_i + a_{i\text{DIF}}\right)}{ a_{i\text{DIF}} } \right) \right) 
        -  b_{i\text{DIF}} \right| & \text{otherwise}.
    \end{cases}
\end{align}

While no formally established classification thresholds exist for the \gls{sa} and \gls{ua} measures, we propose linking these measures for uniform \gls{dif} to the well-established odds-ratio-based effect sizes, specifically the \gls{mh} test with $\Delta_{\text{MH}_i}$ measure \eqref{eq:mh:delta} and the \gls{lr} method with $\Delta_{\text{LR}_i}$ measure \eqref{eq:lr:deltaLR} via the relation
\begin{align*}
    \Delta_{\text{LR}_i} = 2.35\, \beta_{i2} = -2.35\, a_i\, b_{i\text{DIF}},
\end{align*}
using the conventional cut-off values of 1 and 1.5 for its absolute value. Based on this, approximate thresholds for uniform \gls{sa} and \gls{ua} measures can be derived for items with varying discrimination levels: For a~low-discriminating item ($a_i = 0.5$), the thresholds are approximately 0.851 and 1.277. For a~moderately discriminating item ($a_i = 1$), the thresholds are 0.426 and 0.638. Finally, for a highly discriminating item ($a_i = 1.5$), the thresholds drop to 0.284 and 0.426 (Table~\ref{table:ua:thresholds}). 

\begin{table}[H]
    \centering
    \caption{Newly proposed thresholds for the UA measure linked to $\Delta_{\text{LR}_i}$. }\label{table:ua:thresholds}
    \begin{tabular}{l ccc}
        \toprule
        \multirow{3}{*}{Item} & \multicolumn{3}{c}{Classification}\\ \cmidrule(lr){2-4}
        & Negligible & Moderate & Large \\ 
        & (Category A) & (Category B) & (Category C) \\ 
        \midrule
        Low-discriminating & $[0, 0.851)$ & $[0.851, 1.277)$ & $[1.277, \infty)$ \\
        Moderately-discriminating & $[0, 0.426)$ & $[0.426, 0.638)$ & $[0.638, \infty)$ \\
        Highly-discriminating & $[0, 0.284)$ & $[0.284, 0.426)$ & $[0.426, \infty)$ \\
        \bottomrule
    \end{tabular}
\end{table}

To directly account for item discrimination and to enable consistent classification thresholds, we propose \textit{standardized versions} of the area measures, denoted as $\text{SA}_{i}^{\textrm{STD}}$ and $\text{UA}_{i}^{\textrm{STD}}$. These standardized metrics allow the area-based effect sizes to be directly linked to effect-size measures derived from logarithmic odds ratios between groups, facilitating interpretation and comparison in the context of uniform \gls{dif}: 
\begin{align}
     \text{SA}_{i}^{\textrm{STD}} &= \text{SA}_i \cdot a_i = b_{i\text{DIF}} \cdot a_i, \label{eq:irt:es:SAstd} \\
     \text{UA}_{i}^{\textrm{STD}} &= \text{UA}_i \cdot a_i = \vert b_{i\text{DIF}} \vert \cdot a_i. \label{eq:irt:es:UAstd}
\end{align}
For classification of the underlying \gls{dif} effect size, we propose to use the thresholds of 0.426 and 0.638 corresponding to moderately-discriminating items as indicated in Table~\ref{table:ua:thresholds}. In the case of non-uniform \gls{dif}, we consider the same cut-off values for classification. 

The interpretation of the $\text{SA}_{i}^{\textrm{STD}}$ and $\text{UA}_{i}^{\textrm{STD}}$ metrics can be expressed in terms of a shift in the difficulty parameter, as these are equivalent to $a_ib_{i\text{DIF}}$ under uniform \gls{dif} in the \gls{2pl} model. For a moderately-discriminating item ($a_i = 1$), a value of $\text{SA}_{i}^{\textrm{STD}} = 0.426$ corresponds to a difficulty shift of the same magnitude for the focal group. For instance, if the reference group has a 50\% probability of answering this item correctly at an average ability level ($\theta = 0$), the focal group would have a probability of about 39\%. Similarly, a value of  $\text{SA}_{i}^{\textrm{STD}} = 0.638$ corresponds, under the same conditions, to a focal-group probability of about 35\%. These values are consistent with those obtained from the corresponding $\Delta_{\text{MH}}$ cut-offs, providing a coherent interpretation across effect size measures. 

The proposed standardized effect-size measures for uniform \gls{dif} are scale-invariant, meaning that they do not depend on the particular metric used for the latent ability $\theta$. However, in practical applications, the discrimination parameters $a_i$ are unknown and must be estimated from the data, which introduces an additional layer of uncertainty into the effect size estimates.

\paragraph{CSA, CUA: Closed signed and unsigned area measures.}    
By taking the difference over the interval $\left(\theta_1,\, \theta_2\right),$ we get closed versions of the area measures: \gls{csa} and \gls{cua} defined as
\begin{align}
    \text{CSA}_i \label{eq:irt:es:CUA} =&\ \int_{\theta_1}^{\theta_2} \left( F_{i0}  \left( \theta \right) - F_{i1}  \left( \theta \right) \right) \, \mathrm{d}\theta,\\
    \text{CUA}_i \label{eq:irt:es:CSA} =&\ \int_{\theta_1}^{\theta_2} \vert F_{i0}  \left( \theta \right) - F_{i1}  \left( \theta \right)  \vert \, \mathrm{d}\theta. 
\end{align}
Common choices are $\theta_1 = -5$ and $\theta_2 = 5,$ representing a typical range of latent traits. 

\paragraph{WSA, WUA: Weighted area measures.}  
To account for differences in measurement precision, weighting schemes can be incorporated into the calculation of the \gls{sa} \eqref{eq:irt:es:SA} and \gls{ua} \eqref{eq:irt:es:UA}. The resulting effect-size measures \gls{wsa} and \gls{wua} are defined as
\begin{align}
    \text{WSA}_i \label{eq:irt:es:WSA}=& \int_{-\infty}^\infty \left( F_{i0}(\theta) - F_{i1}(\theta) \right) \cdot g(\theta) \, \mathrm{d} \theta, \\
     \text{WUA}_i \label{eq:irt:es:WUA}=& \int_{-\infty}^\infty \left| F_{i0}(\theta) - F_{i1}(\theta) \right| \cdot g(\theta) \, \mathrm{d} \theta,
\end{align}
where $g(\theta)$ denotes the weighting function, e.g., the density of the latent trait in the focal group. The idea that effect-size measures should incorporate information about the latent trait distribution is not new and was already emphasized by \citeA{wainer1993}. In fact, \eqref{eq:irt:es:WSA} with $g$ defined as the density of $\theta$ for the focal group corresponds directly to Wainer’s $\text{T}_i\left( 1 \right)$ index. These measures acknowledge that large differences in functioning do not necessarily indicate problematic bias if only a few individuals are affected. According to \citeA{wainer1993}, the drawback of such refinements is the additional computational complexity they introduce.

Indeed, in the case of \gls{lr}, obtaining a reliable estimate of the integrals in \eqref{eq:irt:es:WSA} and \eqref{eq:irt:es:WUA} is far from straightforward. The standardized total score, which is used as an estimate of $\theta$, typically has only a limited number of discrete values, and therefore will not contribute to a sufficiently accurate estimate of the underlying continuous density without additional assumptions. If one were to approximate the integrals directly using the empirical density of the standardized total score, the result would amount to a very coarse Riemann sum with too few rectangles, leading to a poor approximation of the integral. 

For this reason, we propose to use the standard normal density $\phi \left( \cdot \right)$ as the weighting function, reflecting the common assumption of normally distributed latent traits. The integrals can be approximated numerically by evaluating the item response functions on a fine grid of $\theta$ values and computing weighted sums similarly to \citeA{rudner1977approach}. Specifically, the approximation can be written as
\begin{align}
    \text{WSA}_i  \label{eq:irt:es:WSAap}  &\approx \sum_{k=1}^K \bigl(F_{i0}(\theta_k) - F_{i1}(\theta_k)\bigr)\, \phi(\theta_k)\,\Delta\theta, \\
    \text{WUA}_i \label{eq:irt:es:WUAap} &\approx \sum_{k=1}^K \bigl|F_{i0}(\theta_k) - F_{i1}(\theta_k)\bigr|\, \phi(\theta_k)\,\Delta\theta, 
\end{align}
where $\{\theta_k\}_{k=1}^K$ is a fine grid of evaluation points over the interval $[-5,\,5]$ and $\Delta\theta$ denotes the grid spacing, e.g., 0.01.

In the case of uniform \gls{dif}, \gls{wsa} and \gls{wua} coincide, in the case of non-uniform \gls{dif}, only \gls{wua} should be used due to canceling effects.

\vspace{2ex}

For brevity, effect-size measures are further presented without subscript indices $i$ related to items. Table \ref{tab:effect_sizes:existing} provides an overview of the considered effect-size measures, including their existing classification thresholds and usage restrictions.

\begin{table}[H]
    \centering
    \caption{Summary of existing effect-size measures, their classification thresholds, and corresponding references, along with recommended usage restrictions}
    \resizebox{\textwidth}{!}{%
\begin{tabular}{l cc l r}
    \toprule
    \multirow{2}{*}[-0.25em]{\shortstack[l]{DIF method/ \\ \hspace{1em}Effect-size measure}} &
    \multicolumn{2}{c}{Classification thresholds} &
    \multirow{2}{*}{Reference} &
    \multirow{2}{*}{Usage restrictions} \\
    \cmidrule(lr){2-3}
     & Low & High &  &  \\
    \midrule

    \multicolumn{5}{l}{MH} \\
    \hspace{1em} $|\Delta_{\text{MH}}|$ & 1 & 1.5 & \citeA{holland1985alternate} & Uniform DIF \\ 

    \multicolumn{5}{l}{SIBTEST} \\
    \hspace{1em} $|\hat{\beta}|$ & 0.059 & 0.088 & \citeA{roussos1996simulation} & Uniform DIF \\
                                 & 0.062 & 0.09  & \citeA{weese2022reevaluating} & 2PL model, uniform DIF \\
                                 & 0.069 & 0.102 & \citeA{weese2022reevaluating} & 3PL model, uniform DIF \\

    \multicolumn{5}{l}{Crossing SIBTEST} \\       
    \hspace{1em} $|\hat{\beta}|$ & 0.059 & 0.088 & \citeA{chalmers2018improving} & Non-uniform DIF \\

    \multicolumn{5}{l}{LR} \\
    \hspace{1em} $\Delta_{\text{LR}}$ & 1 & 1.5 & \citeA{monahan2007odds} & Uniform DIF \\ 
    \hspace{1em} $\Delta R^2$ & 0.13 & 0.26 & \citeA{zumbo1997measure} & None \\ 
                              & 0.035 & 0.070 & \citeA{jodoin2001evaluating} & None \\ 
    \hspace{1em} LR-P-DIF & 0.05 & 0.10 & \citeA{monahan2007odds} & None \\ 
    \hspace{1em} LR-STD-P-DIF & 0.05 & 0.10 & \citeA{monahan2007odds} & None \\ 
    \hspace{1em} \makecell[tl]{\gls{sa}, \gls{csa}} & --- & --- & 
    No thresholds available & None \\ 
    \hspace{1em} \makecell[tl]{\gls{ua}, \gls{cua}} & --- & --- & 
    No thresholds available 
    & None\\ 
    \hspace{1em} \makecell[tl]{\gls{wsa}, \gls{wua}} & --- & --- & 
    No thresholds available 
    & Standard normal ability distribution\\ 
    \bottomrule
\end{tabular}
}
\label{tab:effect_sizes:existing}
\end{table}

%-------------------------------------------------------------------
% IMPLEMENTATION
%-------------------------------------------------------------------
\section{IMPLEMENTATION}\label{sec:implementation}

All analyses were conducted in the \texttt{R} software, version 4.2.1 \cite{R}. The implementation of the proposed effect-size measures and classification guidelines was prepared using the \texttt{difR} package \cite{difR}, version 6.1.0 and the \texttt{difNLR} package, version 1.5.2-2 \cite{hladka2020difnlr}. An interactive implementation is also provided through the \texttt{ShinyItemAnalysis} application, version 1.6.0  \cite{martinkova2018shinyitemanalysis, martinkova2026SIAmodules}, which supports a wide range of psychometric analyses of multi-item measurements without programming. Sample \texttt{R} code implementing the proposed methods on real data examples, together with an illustration of the interactive workflow, is available in the Online Supplementary Material.

Data generation in simulations and \gls{dif} detection were performed via the \texttt{difR} and \texttt{difNLR} packages. A graphical representation of the results was created with the \texttt{ggplot2} package, version 4.0.0 \cite{ggplot2} and the \texttt{viridis} package, version 0.6.5 \cite{garnier2024virids}. Tables were created using the \texttt{xtable} package, version 1.8-4 \cite{dahl2019xtable}. Data manipulation was done using the \texttt{dplyr} package, version 1.1.4 \cite{wickham2023dplyr} and the \texttt{tidyr}, version 1.3.1. \cite{wickham2024tidyr}.

%-------------------------------------------------------------------
% SIMULATION STUDY
%-------------------------------------------------------------------
\section{SIMULATION STUDY}\label{sec:simulation}

A simulation study was conducted to evaluate the performance of existing classification guidelines for selected effect-size measures and to assess their properties. The analysis focused on how these estimates are influenced by design factors, such as sample size, and on their effectiveness in distinguishing between varying levels of the underlying magnitude of \gls{ua} in \eqref{UA}. Based on the findings, we proposed updated cut-off values to enhance comparability across different \gls{dif} detection methods and related effect-size measures. Finally, we introduced new classification thresholds for area-based effect sizes, aiming to support more consistent and robust applications in practice. 

%-------------------------------------------------------------------
\subsection{Simulation design}

%-----------------------
\subsubsection{Data generation} 

The data for the simulation study were generated by manipulating six design factors: (a) sample size, (b) test length, (c) proportion of \gls{dif} items, (d) ability distribution for the focal group, (e)magnitude of the underlying \gls{dif} effect, and (f) type of \gls{dif}. 

Sample sizes of 100, 200, 500, 1{,}000, 2{,}000, 5{,}000, and 10{,}000 were selected for both groups, representing a range from small sample sizes to large-scale assessments. Each simulated test consisted of either 20 or 40 items, with 5\% or 10\% designated as \gls{dif} items. 

The ability distribution of the reference group was simulated using a standard normal distribution. For the focal group, three scenarios were considered: In the first scenario, the ability levels were equivalent to the levels of the reference group, drawn from the standard normal distribution; in the second case, the ability levels for the focal group were drawn from a shifted normal distribution with a mean of $-0.5$ and a standard deviation of 1, following \citeA{jodoin2001evaluating}. Additionally, we included the third case, where the ability levels for the focal group were drawn from a shifted normal distribution with a mean of $0.5$ and a standard deviation of 1, to directly assess the symmetry of the results with respect to the direction of the ability shift. 

Item responses were generated under a true underlying \gls{2pl} \gls{irt} model. In all scenarios, parameters for non-\gls{dif} items were generated randomly, with discrimination parameters drawn from $a \sim \mathcal{N} \left(1,\, 0.2\right)$ and difficulty parameters from $b \sim \mathcal{N} \left(0,\, 0.25\right)$, reflecting realistic ranges observed in practice. This approach ensures broad coverage of plausible item characteristics and allows for systematic investigation under controlled conditions. The same parameters were used for reference and focal groups, resulting in no underlying \gls{dif}. 

Additionally, a \gls{4pl} was considered for generating non-\gls{dif} items as part of a sensitivity analysis, with lowe and upper asymptotes drawn from $c \sim \text{Unif} \left(0,\, 0.25\right)$ and $d \sim \text{Unif} \left(0.75,\, 1\right)$, respectively, to better reflect realistic testing condition. All item parameters for non-\gls{dif} items are summarized in Supplementary Table \ref{app:tab:nonDIFparameters}. For the \gls{2pl} \gls{irt} model, asymptotes were fixed at $c = 0$ and $d = 1$.

Magnitude of the underlying \gls{dif} effect was defined by the \gls{ua} between theoretical \glspl{icc} following \citeA{swaminathan1990detecting}, \citeA{narayanon1996identification}, and \citeA{drabinova2017detection}. For each selected level of the area-based effect size (0.2, 0.4, 0.6, 0.8, 1.0, and 1.2), four \gls{dif} items were generated using a \gls{2pl} \gls{irt} model. The choice of effect size levels was inspired by \citeA{narayanon1996identification}; however, an additional small effect (0.2) and large effects (1 and 1.2) were included to evaluate performance in more extreme scenarios and to enable more precise determination of classification thresholds. 

For uniform \gls{dif} items, the discrimination parameter $a$ was fixed at either 1 or 1.5 for both groups. The difficulty parameter $b$ was set to 0, $-0.5$, and 0.5 in the reference group for $a = 1$, and to 0 for $a = 1.5$. The focal-group difficulty was then adjusted to achieve the desired levels of the underlying area-based effect size, resulting in four uniform \gls{dif} items, see Equations \eqref{SA} and \eqref{UA}.

Similarly, for non-uniform \gls{dif} items, the discrimination parameter $a$ was fixed at either 0.5 or 0.4 for the reference group. The difficulty parameter $b$ was set to 0, $-0.5$, and 0.5 for $a = 0.5$ and to 0 for $a = 0.4$, in both groups. The focal-group discrimination was then adjusted to achieve the desired levels of the underlying area-based effect size, resulting in four non-uniform \gls{dif} items. An overview for both uniform and non-uniform \gls{dif} item specifications is provided in Supplementary Table \ref{app:tab:DIFparameters}. For each simulated dataset, the required number of \gls{dif} items was selected sequentially from the beginning of Supplementary Table~\ref{app:tab:DIFparameters}. The remaining items were filled with non-\gls{dif} items drawn sequentially from Supplementary Table~\ref{app:tab:nonDIFparameters} until the desired total number of items was reached. Results of the sensitivity analysis can be found in Table \ref{tab:sensitivity_uniform} and Table \ref{tab:sensitivity_nonuniform}.

In total, the simulation study includes 7 (sample sizes) $\times$ 2 (test lengths) $\times$ 2 (proportions of \gls{dif} items) $\times$ 3 (latent ability distributions for the focal group) $\times$ 6 (underlying \gls{dif} effect size values) $\times$ 2 (types of \gls{dif}) $=$ 1{,}008 different settings. Each simulation setting was generated 1{,}000 times.

%-----------------------
\subsubsection{Statistical analysis} 

For uniform \gls{dif}, we conducted the \gls{mh} test, the \gls{sibtest} method, and the \gls{lr} approach with the likelihood ratio test (see Table~\ref{tab:tests}, second row), computing their respective effect-size measures and applying classification thresholds: $\vert \Delta_{\text{MH}} \vert$ \eqref{eq:mh:delta} with \eqref{eq:mh:es}, $\vert \hat{\beta} \vert$  \eqref{eq:sibtest:hatbeta} with \eqref{eq:sibtest:hatbeta:thresholds}, $\Delta R^2$ \eqref{eq:lr:R2diff}, $\Delta_{\text{LR}}$ \eqref{eq:lr:deltaLR}, $\text{UA}^{\textrm{STD}}$ \eqref{eq:irt:es:UAstd}, and $p$ metrics  LR-P-DIF \eqref{eq:lr:LRPDIF} and LR-STD-P-DIF \eqref{eq:lr:LRSTDPDIF} considering total, uniform, reference, and Cochran's weights using \eqref{eq:lr:pdif}. For non-uniform \gls{dif}, we applied the \gls{sibtest} method and the \gls{lr} model with the likelihood ratio test (see Table~\ref{tab:tests}, third row) and the corresponding cut-off values for classification: $\vert \hat{\beta} \vert$  \eqref{eq:sibtest:hatbeta:nunif} with \eqref{eq:sibtest:hatbeta:thresholds} and $\Delta R^2$ \eqref{eq:lr:R2diff}. Additionally, for both types of \gls{dif}, we computed several area-based effect-size measures using the \gls{lr} model: \gls{sa} \eqref{SA}, \gls{ua} \eqref{UA}, \gls{csa} \eqref{eq:irt:es:CSA}, \gls{cua} \eqref{eq:irt:es:CUA}, and \gls{wua} \eqref{eq:irt:es:WUAap} (with step widths of 0.01). All statistical analyses were performed at a 5\% significance level. 

To assess the efficiency of existing guidelines, we calculated the proportions of item classifications (non-significant, negligible, moderate, and large) using established cut-off values. These calculations were conducted separately for subgroups defined by the magnitude of the underlying \gls{dif} (as simulated by the magnitude of \gls{ua}) and sample size. The analysis included both \gls{dif} and non-\gls{dif} items: for non-\gls{dif} items, to determine the proportion of false positives (non-\gls{dif} items incorrectly identified as functioning differentially), and for \gls{dif} items, to assess the alignment between existing classification and empirical results.

Further, we performed a linear regression analysis to estimate the relationship between underlying and estimated \gls{ua}, and, for underlying uniform \gls{dif}, also between the underlying and estimated $\text{UA}^{\textrm{STD}}$. The impact of the design factors -- sample size, test length, the proportion of \gls{dif} items, and a distributional shift in the focal group's ability -- was investigated using linear regression models. The sample size was treated as a continuous variable (with a linear effect), while the remaining factors were treated as categorical variables. The model included two-way interactions.

To link the different effect-size measures, we fitted linear regression models: for uniform \gls{dif}, other non-area-based effect sizes were regressed on $|\Delta_{\text{MH}}|$, and for non-uniform \gls{dif}, other effect sizes were regressed on \gls{ua}. These models were used to derive cut-off values aligning with cut-off values 1 and 1.5 for $\vert \Delta_{\text{MH}} \vert$ and 0.426 and 0.638 for \gls{ua}. Model selection was based on likelihood-ratio tests. Linear dependency models sufficed for most effect-size measures, except for $\Delta R^2$, which required modeling a cubic relationship. The correlation between effect-size measures was estimated by the Pearson correlation coefficient.

Lastly, we compared the proportions of correctly classified cases between the existing and newly proposed classification guidelines. The correct classification was determined based on the underlying value of \gls{ua} using the theoretically-derived thresholds of 0.426 and 0.638 for \gls{ua}.

%-------------------------------------------------------------------
\subsection{Simulation results}\label{sec:results}

%-----------------------
\subsubsection{Uniform DIF}

\begin{figure}[h!]
    \centering
    \caption{Uniform DIF effect size classification for different effect-size measures (rows) and underlying effect-size magnitudes (columns) based on a sample size (inner columns). }
    \includegraphics[width=1.0\textwidth]{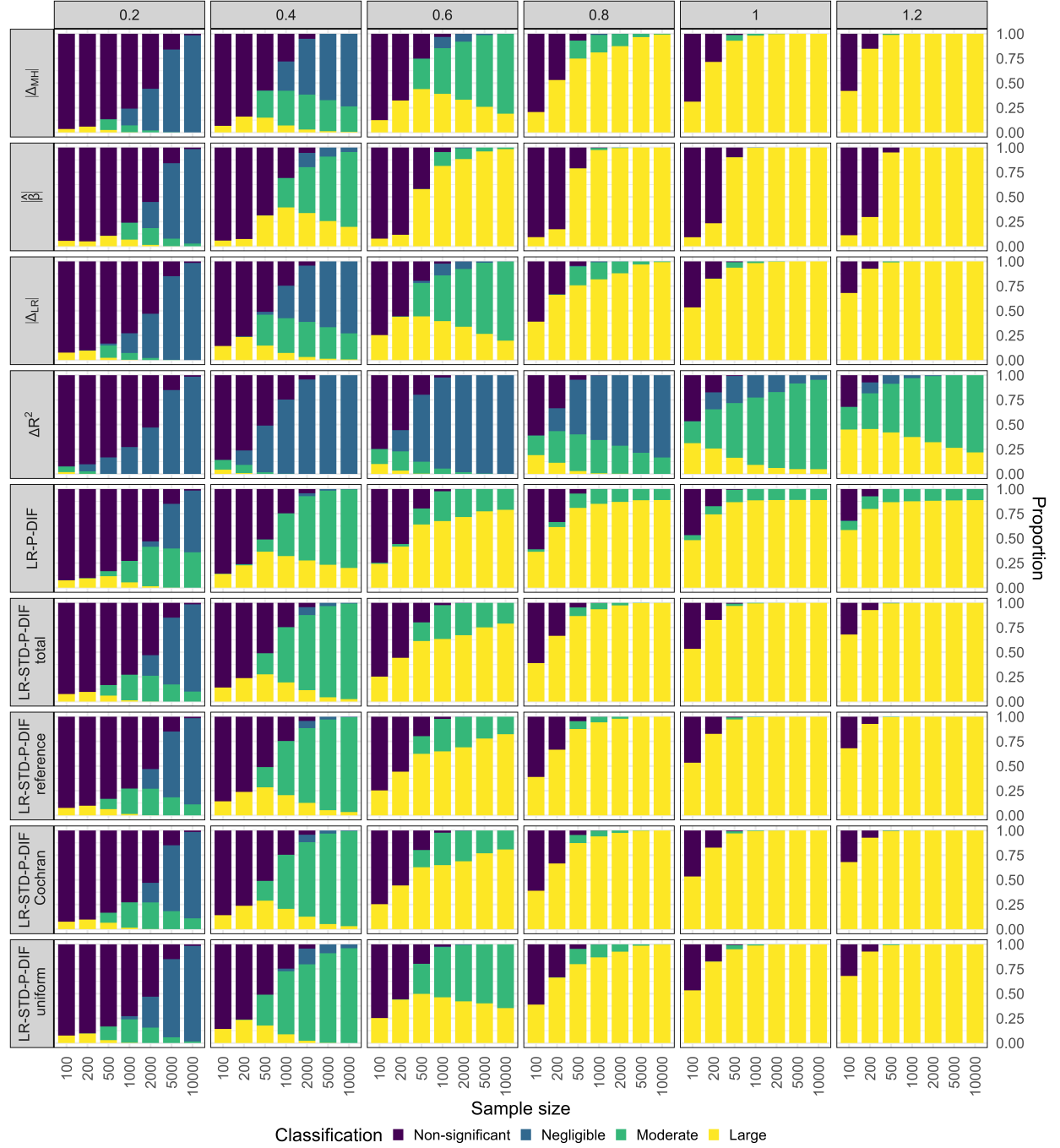}
    \label{fig:uniform_proportion_TP}
\end{figure}

\paragraph{Current classification: \gls{dif} items. }
For the underlying \gls{ua} of 0.2, effect sizes across all measures and sample sizes were predominantly classified as non-significant or negligible according to existing classification guidelines (see Figure~\ref{fig:uniform_proportion_TP}). However, the LR-STD-P-DIF (excluding uniform weights) and LR-P-DIF $p$ metrics showed substantial rates of misclassifications, with a~notable proportion of effect sizes classified as moderate, especially at larger sample sizes ($N \geq 1{,}000$; ranging from 10\% to 40\%; see the first column and rows 5--8 in Figure~\ref{fig:uniform_proportion_TP}).

For \gls{ua} of 0.4, most effect sizes were classified as non-significant at smaller sample sizes ($N \leq$ 500). For larger sample sizes, $|\Delta_{\text{MH}}|$ and $\Delta_{\text{LR}}$ tended to yield classifications of negligible to moderate, while the \gls{sibtest}'s $|\hat{\beta}|$, LR-STD-P-DIF, and LR-P-DIF showed more frequent moderate to large classification. In contrast, the effect sizes of $\Delta R^2$ were mainly negligible. For the underlying \gls{ua} of 0.6 and sample size $N \geq$ 500, most effect sizes were classified as moderate or large across all methods, except $\Delta R^2$ where the effects remained negligible. \gls{sibtest}'s $|\hat{\beta}|$, LR-STD-P-DIF, and LR-P-DIF showed an increased proportion of classified large effect sizes with increasing sample size.  

For \gls{ua} values of 0.8, 1, and 1.2, the effect sizes were predominantly classified as large for all measures across all sample sizes, except for $\Delta R^2$. In the case of $\Delta R^2$, classifications remained mostly non-significant (at smaller sample sizes) or negligible (at larger sample sizes) for $\text{UA} = 0.8$ (ranging from 57\% to 83\%) and shifted to moderate for \gls{ua} values of 1 and 1.2 (ranging from 22\% to 91\%; see the fourth row in Figure~\ref{fig:uniform_proportion_TP}). 

\paragraph{Current classification: non-\gls{dif} items.}
The use of effect-size measures helped reduce false positives, as the majority of effect sizes for non-\gls{dif} items (i.e., those with an underlying \gls{ua} of 0) were classified as negligible according to the existing classification guidelines. However, the proportion of falsely detected \gls{dif} items slightly increased with both sample size and underlying \gls{ua} of \gls{dif} items within the same dataset. This trend was consistent across all effect-size measures; hence, only summarized results are presented.

\begin{figure}[H]
   \centering
   \caption{Uniform DIF effect size classification for non-DIF items for all effect-size measures and the underlying DIF effect sizes (columns) based on a sample size (inner columns).}
   \includegraphics[width=\textwidth]{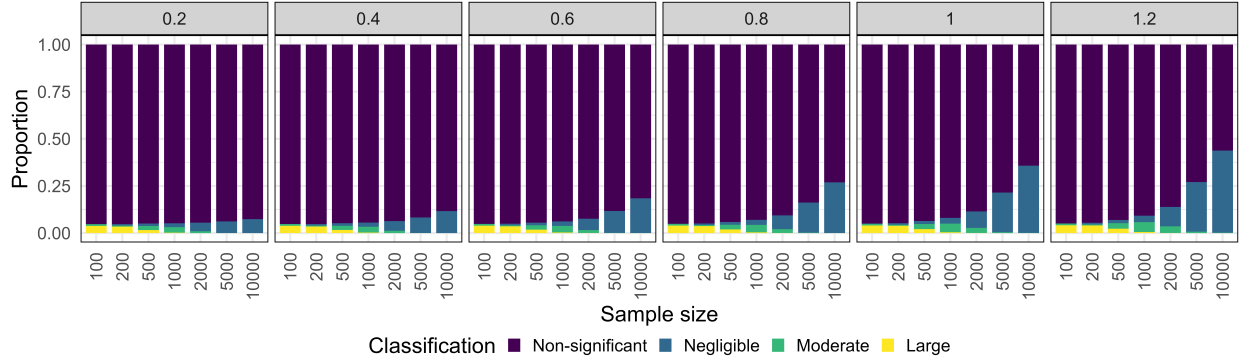}
   \label{fig:uniform_proportion_FP_alternative}
   \vspace{-2em}
\end{figure}

For small sample sizes ($N \leq$ 200) and across all \gls{ua} levels, 95\% of all falsely detected items were classified as non-significant, 1\% as negligible or moderate, and 4\% as large. In contrast, for larger sample sizes, the rate of false positives increased, especially with higher values of \gls{ua}. For example, at $N = 10{,}000$ and $\text{UA} = 0.8$, 64\% of all non-\gls{dif} items were classified as non-significant and 36\% as negligible. For the same sample size and a \gls{ua} of 1.2, 56\% were non-significant and 44\% negligible (Figure \ref{fig:uniform_proportion_FP_alternative}). 

\paragraph{Underlying and estimated area measure.}
The estimated \gls{ua} measures were slightly overestimated for small underlying \gls{ua} (0.22 for 0.2) and slightly underestimated for moderate \gls{dif} (0.38 for 0.4; 0.55 for 0.6) and larger effects (0.72 for 0.8; 0.89 for 1; 1.06 for 1.2). While the effects of sample size, test length, the proportion of \gls{dif} items, and distributional shifts in focal group ability on \gls{ua} were statistically significant due to a large number of observations (over 14 million; p-values < 0.0001), their practical impact on predictions was minimal. Overall, \gls{ua} estimates appeared stable with respect to these external factors. Similar consistency results, i.e., dependence only on the underlying value of \gls{ua}, were observed for the $|\Delta_{\text{MH}}|$ statistic. 

Similarly, for the standardized version $\text{UA}^{\textrm{STD}}$, the estimates tended to be slightly overestimated for small underlying values of $\text{UA}^{\textrm{STD}}$ (0.24 for 0.2; 0.33 for 0.3), accurate for moderate values (0.42 for 0.4; 0.6 for 0.6) and slightly underestimated for larger effects (0.78 for 0.8; 0.96 for 1, 1.14 for 1.2, 1.41 for 1.5, 1.68 for 1.8). As with \gls{ua}, the dependency on design factors had a minimal impact on predictions.

%-----------------------
\subsubsection{Non-uniform DIF}

\begin{figure}[hbt]
    \caption{Non-uniform DIF effect size classification for the crossing SIBTEST's $|\hat{\beta}|$ and LR's $\Delta R^2$ (rows) and the underlying DIF effect sizes (columns) for DIF items based on a sample size (inner columns).}
    \centering
    \includegraphics[width=\textwidth]{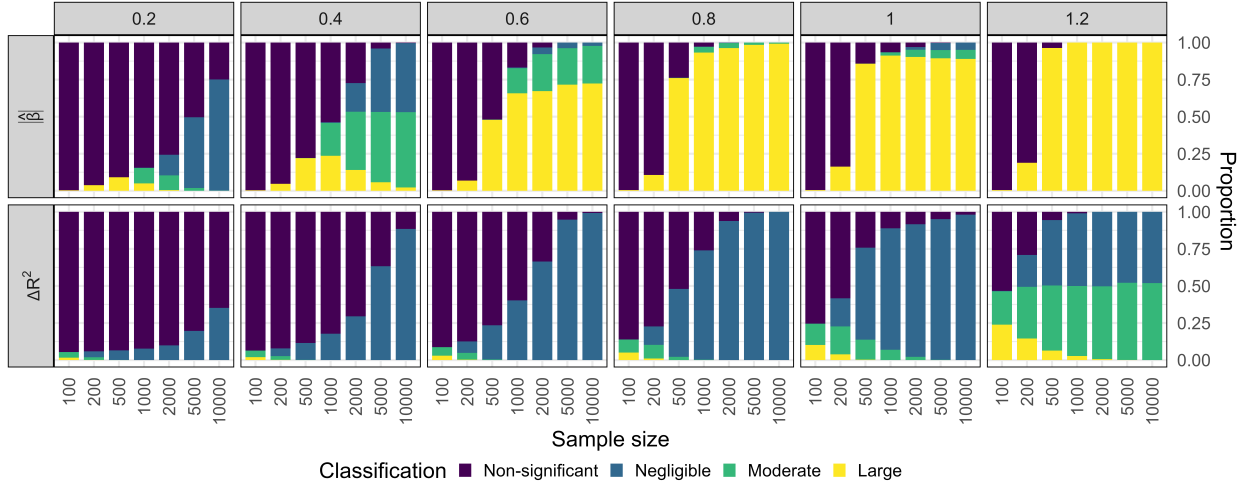}
    \label{fig:nonuniform_proportion_TP}
    \vspace{-2em}
\end{figure}

\paragraph{Current classification: \gls{dif} items.}
For the underlying \gls{ua} values between 0.2 and 0.6, the majority of effect sizes for both measures $|\hat{\beta}|$ (from \gls{sibtest}) and $\Delta R^2$ (from \gls{lr}) were either non-significant or classified as negligible. At higher levels of \gls{ua}, both effect-size measures demonstrated sensitivity to sample size but in different ways (Figure~\ref{fig:nonuniform_proportion_TP}). 

Using \gls{sibtest}'s $|\hat{\beta}|$, items with a \gls{ua} of 0.8 were mostly classified as non-significant at smaller sample sizes ($N \leq 200$). However, as sample size increased ($N \geq 500$), a growing proportion of \gls{dif} items was classified as having a large effect, ranging from 76\% to 96\%. Similarly for \gls{ua} values of 1 and 1.2, items were non-significant at small sample sizes ($N \leq 200$), but the proportion classified as having large \gls{dif} increased with sample size, ranging from 86\% to 95\% for $\text{UA} = 1$ and from 96\% to 100\% for $\text{UA} = 1.2$ (see the first row in Figure~\ref{fig:nonuniform_proportion_TP}). 

The $\Delta R^2$ values showed a different pattern. For a \gls{ua} of 0.8, items were mostly non-significant at smaller sample sizes ($N \leq 500$) and typically classified as negligible at larger sample sizes ($N \geq 1{,}000$). A similar pattern held for $\text{UA} = 1$. At $\text{UA} = 1.2$, a larger proportion of \gls{dif} items was classified as moderate (ranging from 22\% to 52\%), with only a small fraction reaching the large \gls{dif} effect size classification (up to 24\%). However, this trend was not monotonic (see the second row in Figure~\ref{fig:nonuniform_proportion_TP}). 

\begin{figure}[hbt]
    \caption{Non-uniform DIF effect size classification for the crossing SIBTEST's $|\hat{\beta}|$ and LR's $\Delta R^2$ (rows) and the underlying DIF effect sizes (columns) for non-DIF items based on a sample size (inner columns).}
    \centering
    \includegraphics[width=1\textwidth]{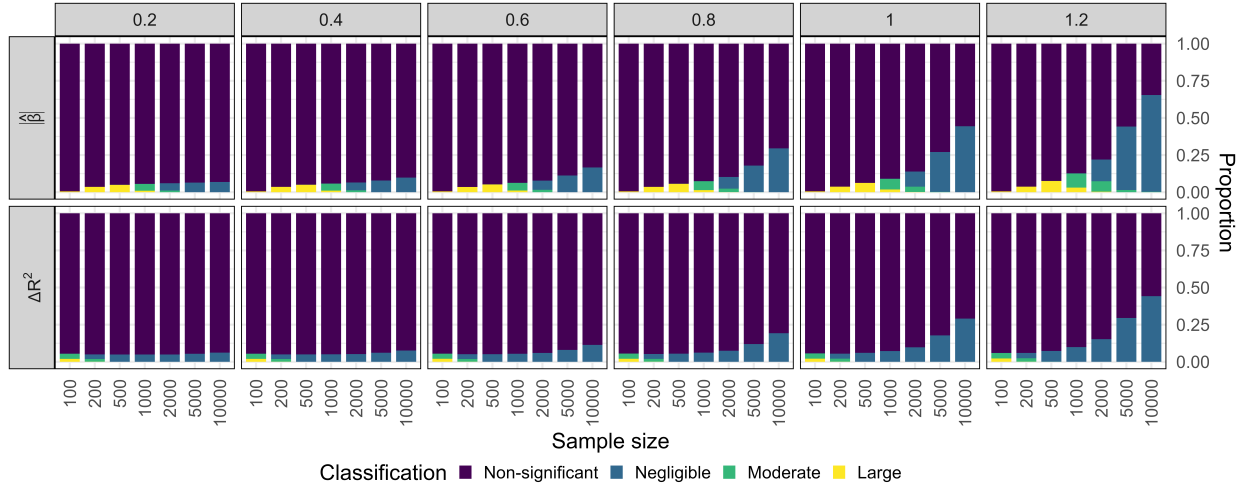}
     \label{fig:nonuniform_proportion_FP}
     \vspace{-2em}
\end{figure}

\paragraph{Current classification: non-\gls{dif} items. }
As for the uniform \gls{dif}, the use of effect-size measures helped to reduce the proportions of detected false positives. The crossing \gls{sibtest} method performed well in controlling Type I error, although the nominal level was slightly exceeded, primarily at larger sample sizes ($N \geq 1{,}000$). Using \gls{sibtest}'s $|\hat{\beta}|$, non-\gls{dif} items that were falsely detected tended to be classified as having a large effect at smaller sample sizes ($N \leq$ 500), and as having a negligible effect for larger sample sizes (see the first row in Figure~\ref{fig:nonuniform_proportion_FP}). In contrast, with the \gls{lr} method, rejection rates exceeded the nominal level only at very large sample sizes ($N \geq 5{,}000$) and higher underlying \gls{ua} (\gls{ua} $\geq$ 0.8). However, even in these cases, the falsely identified items were mainly classified as having a negligible \gls{dif} effect size (see the second row in Figure~\ref{fig:nonuniform_proportion_FP}). 

\paragraph{Underlying and estimated area measure.} For non-uniform \gls{dif}, \gls{ua} was highly variable for small sample sizes, i.e. $N \leq 200.$ For $N \geq 500,$ the estimates of \gls{ua} were slightly overestimated (0.23 for 0.2; 0.45 for 0.4; 0.68 for 0.6; 0.9 for 0.8; 1.13 for 1; 1.35 for 1.2). Again, the dependency on other design factors had no practical impact on predicted values.

%-------------------------------------------------------------------
\subsection{Newly proposed cut-off values}

%-----------------------
\subsubsection{Uniform DIF}

\begin{figure}[h]
    \centering
    \caption{Dependencies of effect-size measures on sample size with classification cut-offs indicated by red dashed lines.}
    \begin{subfigure}[b]{0.495\textwidth}
        \centering
        \caption{$|\Delta_{\text{MH}}|$ with cut-offs of 1 and 1.5.}
        \includegraphics[width=\textwidth]{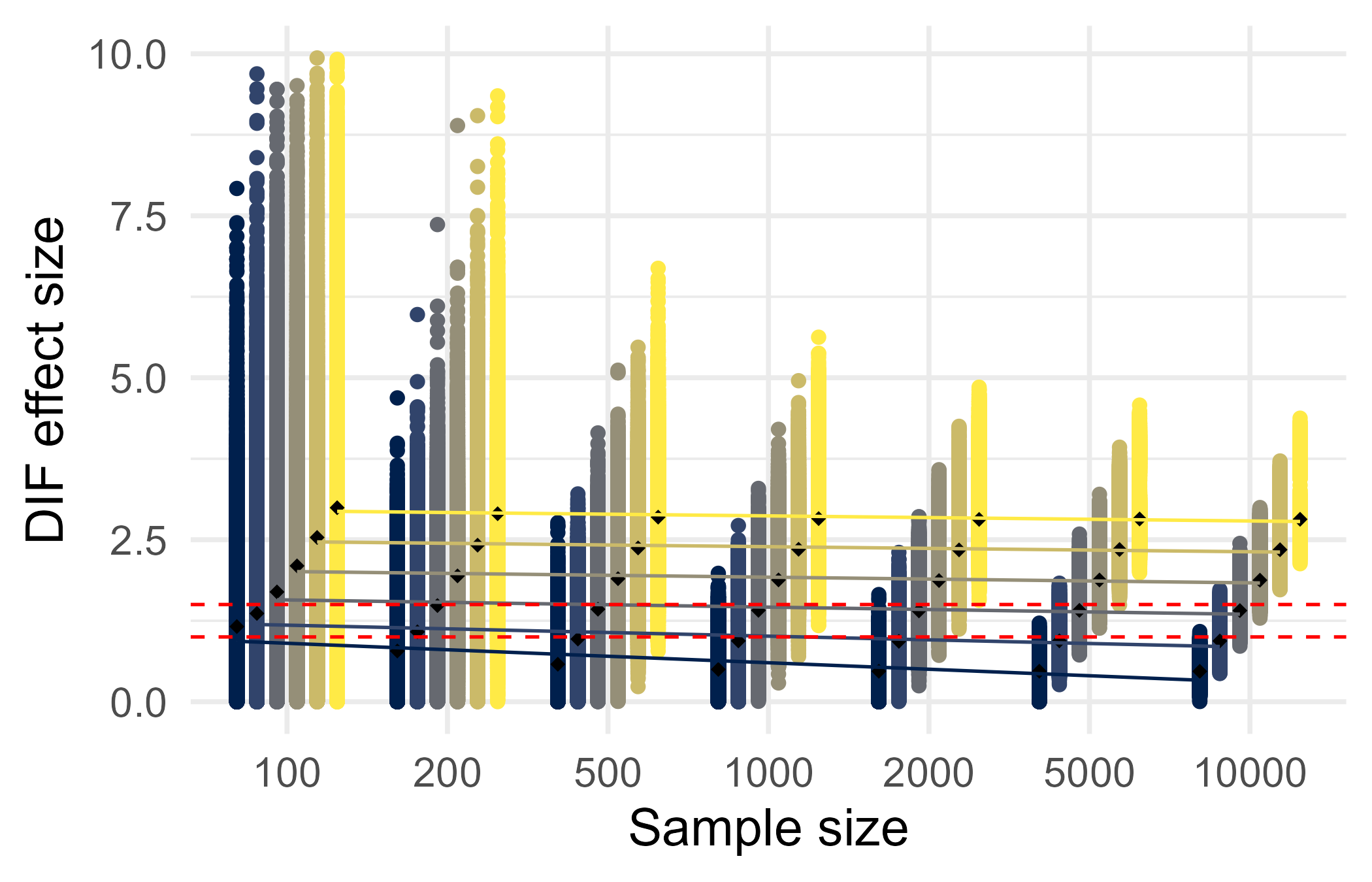}
        \label{fig:es1}
    \end{subfigure}
    \hfill
    \begin{subfigure}[b]{0.495\textwidth}
        \centering
        \caption{SIBTEST $|\hat{\beta}|$ with cut-offs of 0.059 and 0.088.}
        \includegraphics[width=\textwidth]{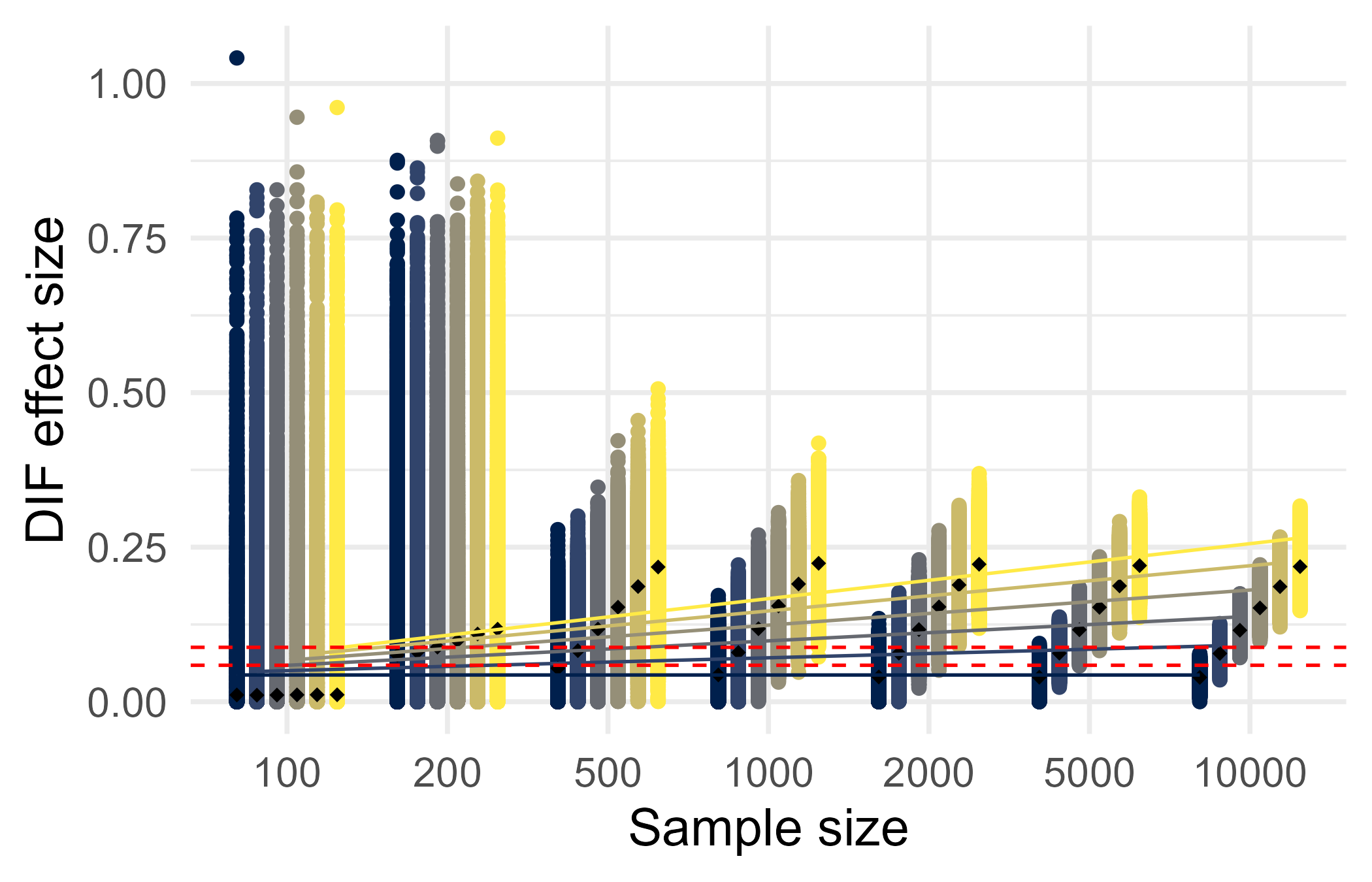}
        \label{fig:es2}
    \end{subfigure}
    
    \vspace{-0.3cm}

    \begin{subfigure}[b]{0.495\textwidth}
        \centering
        \caption{$\Delta R^2$ with cut-offs of 0.035 and 0.07.}
        \includegraphics[width=\textwidth]{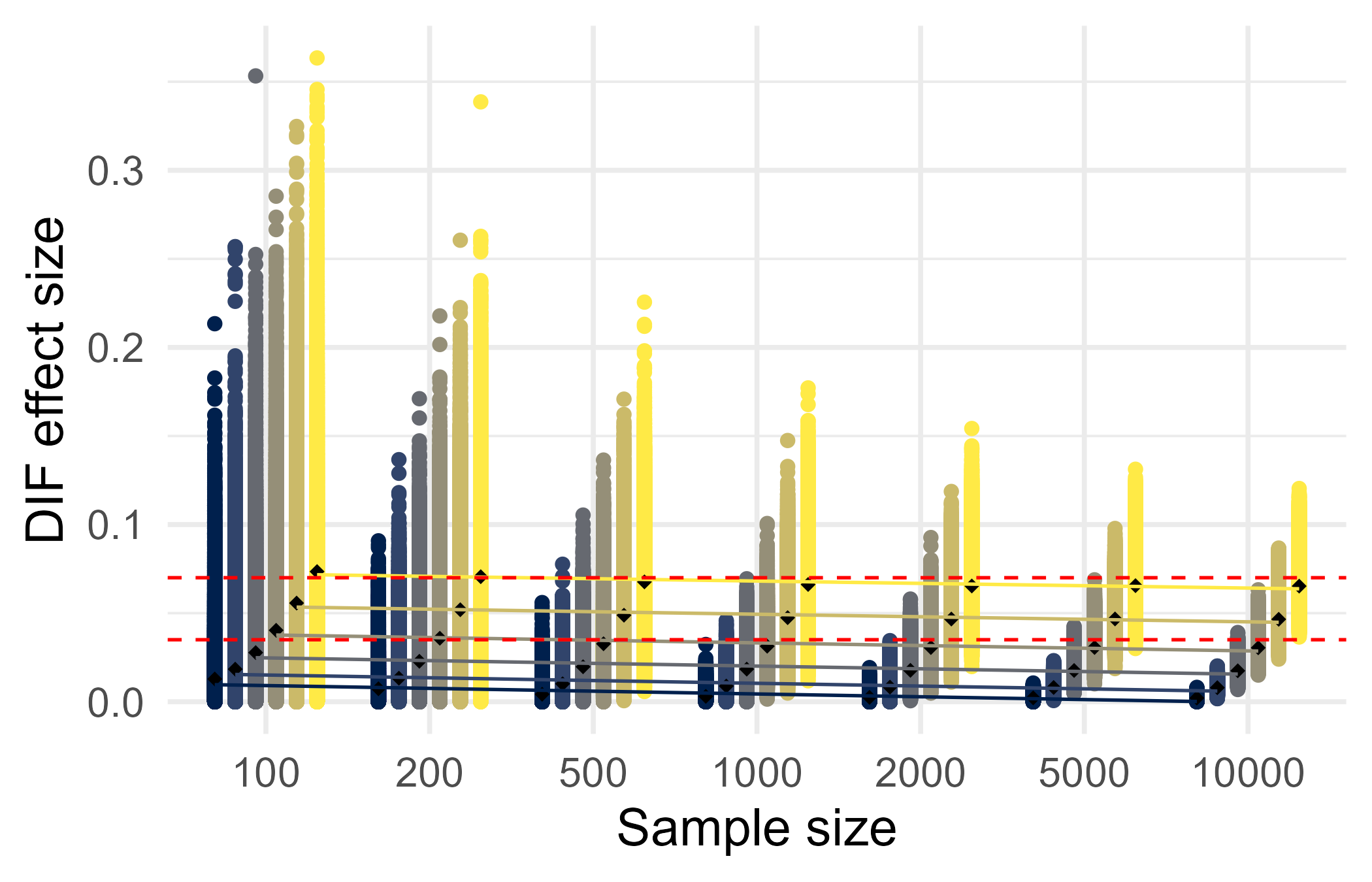}
        \label{fig:es3}
    \end{subfigure}
    \hfill
    \begin{subfigure}[b]{0.495\textwidth}
        \centering
        \caption{LR-P-DIF with cut-offs of 0.05 and 0.1.}
        \includegraphics[width=\textwidth]{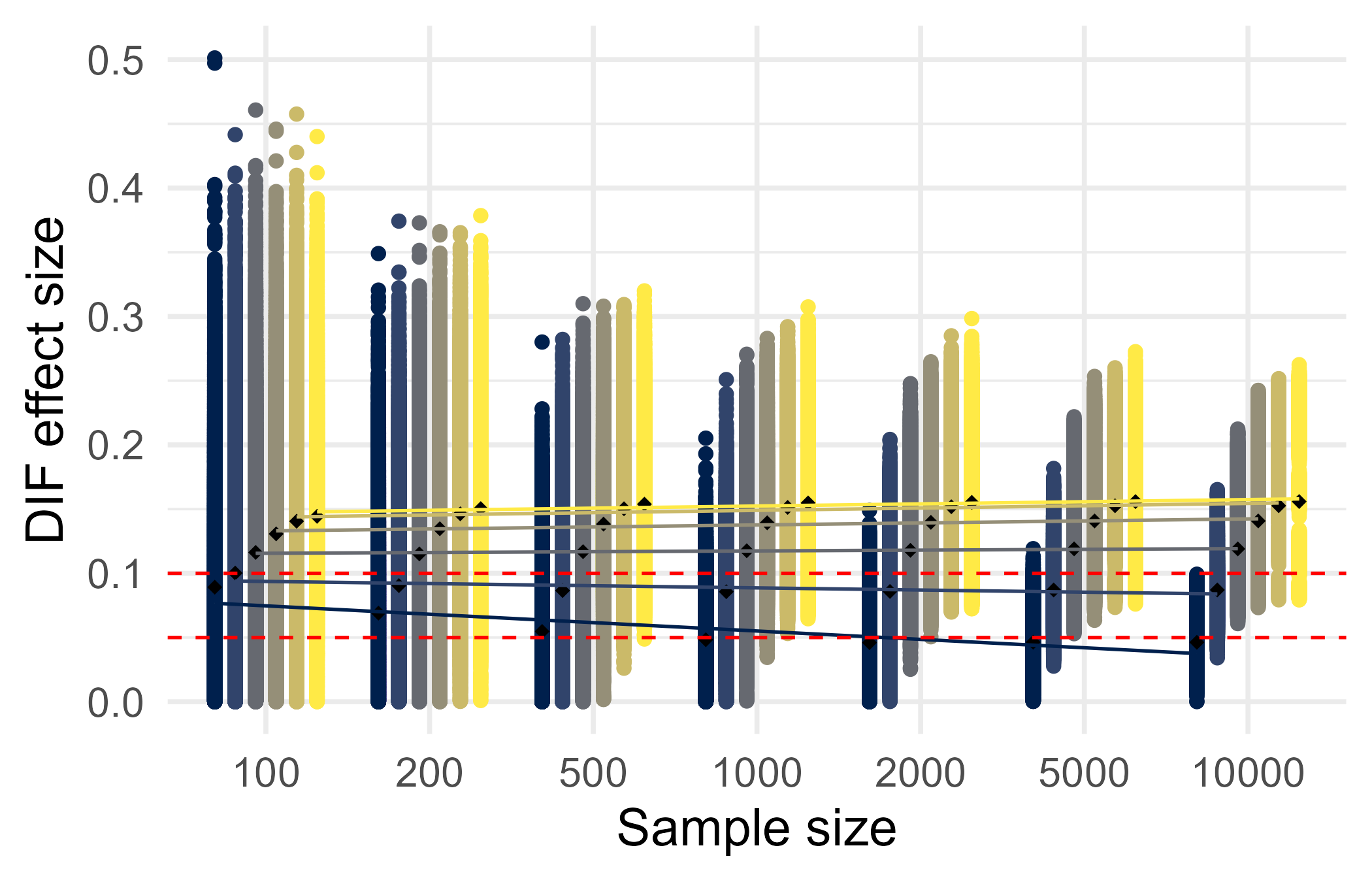}
        \label{fig:es4}
    \end{subfigure}
    
    \vspace{-0.4cm} 
    
    \includegraphics[width=0.75\textwidth]{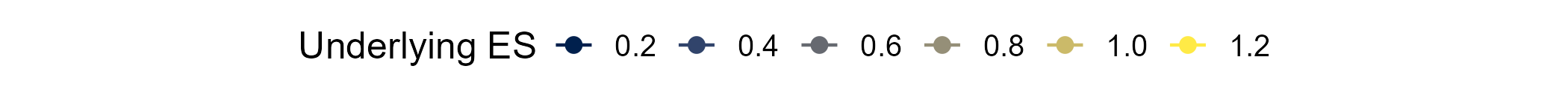}  
    
    \label{fig:esplots}
\end{figure}

We propose updated classification thresholds for multiple \gls{dif} effect-size measures. For the \gls{mh} measure $|\Delta_{\text{MH}}|$, we retained cut-off values of 1 (negligible/moderate) and 1.5 (moderate/large). This choice was justified by the measure's clear distinction between effect size levels and its minimal dependency on sample size, with only a minor decline at the lowest underlying \gls{ua} level of 0.2 (Figure \ref{fig:es1}). Furthermore, as the earliest and most widely adopted effect-size metric, $|\Delta_{\text{MH}}|$ provides a natural reference point for linking to area-based measures through $\text{UA}^{\textrm{STD}}$~\eqref{eq:irt:es:UAstd}.

Using $|\Delta_{\text{MH}}|$ as the anchor, we established corresponding thresholds for other \gls{dif} effect-size measures. For \gls{sibtest}'s $|\hat{\beta}|$, exploratory analyses revealed problematic behavior at small sample sizes ($N \leq$ 200), including inflated variances (Figure \ref{fig:es2}) and weak associations with $|\Delta_{\text{MH}}|$ ($\hat{\rho} = 0.16$ at $N = 100$; $\hat{\rho} = 0.33$ at $N = 200$). At moderate to large sample sizes, however, correlations were strong and linear ($\hat{\rho} =$ 0.88--0.98). We therefore excluded results from $N \leq 200$ when determining thresholds and recommend cut-off values of 0.085 (negligible/moderate) and 0.122 (moderate/large).

In contrast, \gls{lr}'s $\Delta R^2$ measure does not show a trend with sample size (Figure~\ref{fig:es3}). Based on a cubic regression model on $|\Delta_{\text{MH}}|$, we propose thresholds of 0.01 (negligible/moderate) and 0.021 (moderate/large). Among the $p$ metrics, the non-standardized LR-P-DIF performed poorly, failing to effectively distinguish between underlying \gls{ua} values in the range of 0.8 to 1.2 (Figure \ref{fig:es4}) and showing sensitivity to shifts in the focal group's ability distribution. In contrast, its standardized variant, LR-P-STD-DIF, mitigated these issues. Using linear regression on $|\Delta_{\text{MH}}|$, we recommend thresholds of 0.084 (negligible/moderate) and 0.116 (moderate/large), for Cochran's, reference, and total weights, which yielded comparable results. Uniform weights yielded slightly lower thresholds of 0.075 and 0.116. The $\Delta_{\text{LR}}$ measure demonstrated a strong correlation with $|\Delta_{\text{MH}}|$ ($\hat{\rho} = 0.91$), with the strength of the linear association increasing alongside sample size. Consequently, a linear regression analysis supported the same cut-off values as those proposed for $|\Delta_{\text{MH}}|$: 1 for moderate and 1.5 for large effect sizes. 

For area-based effect-size measures, differences across variations were minimal. As expected, \gls{sa} and \gls{ua} are equivalent for uniform \gls{dif}, while \gls{ua} and \gls{cua} showed almost identical behaviour ($\hat{\rho} = 0.99$). We therefore propose unified thresholds of 0.426 (negligible/moderate) and 0.638 (moderate/large) for \gls{sa}, \gls{ua}, and \gls{cua}. \gls{wua} demonstrated sensitivity to shifts in the focal group's ability distribution with no practical impact on predictions, yet retained a strong linear relationship with $|\Delta_{\text{MH}}|$ ($\hat{\rho} = 0.94$), yielding regression-based thresholds of 0.085 (negligible/moderate) and 0.117 (moderate/large). Finally, $\text{UA}^{\textrm{STD}}$ was highly correlated with $|\Delta_{\text{MH}}|$ ($\hat{\rho} = 0.95$), further supporting its role as a bridge between odds-ratio-based and area-based effect sizes.

A complete summary of all proposed cut-off values, including recommended usage restrictions, is provided in Table \ref{tab:cutoffs:uniform}. 

\begin{table}[ht]
    \centering
    \caption{Newly proposed classification thresholds for effect-size measures used in detecting uniform DIF, along with recommended usage restrictions}
    \resizebox{\textwidth}{!}{%
\begin{tabular}{l cc l}
    \toprule
    \multirow{2}{*}[-0.25em]{\shortstack[l]{DIF method/ \\ \hspace{1em}Effect-size measure}} & 
    \multicolumn{2}{c}{Classification thresholds} &
    \multirow{2}{*}{Usage restrictions} \\
    \cmidrule(lr){2-3}
     & Low & High \\
    \midrule

    \multicolumn{4}{l}{MH} \\
    \hspace{1em} $|\Delta_{\text{MH}}|$ & 1 & 1.5 & None \\ 

    \multicolumn{4}{l}{SIBTEST} \\
    \hspace{1em} $|\hat{\beta}|$ & 0.085 & 0.122 & $N \geq 500$ \\ 

    \multicolumn{4}{l}{LR} \\
    \hspace{1em} $\Delta_{\text{LR}}$ & 1 & 1.5 & None \\ 
    \hspace{1em} $\Delta R^2$ & 0.010 & 0.021 & None \\  
    \hspace{1em} LR-P-DIF & 0.092 & 0.122 & Standard normal ability distribution \\ 
    \hspace{1em} \makecell[tl]{LR-STD-P-DIF\\ (total, reference, and\\ Cochran's weights)} & 0.084 & 0.116 & None \\ 
    \hspace{1em} \makecell[tl]{LR-STD-P-DIF\\ (uniform weights)} & 0.075 & 0.104 & None \\ 
    \hspace{1em} \makecell[tl]{\gls{sa}, \gls{ua}}%\gls{csa}} 
    & 0.426 & 0.638 & None \\ 
    \hspace{1em} \makecell[tl]{\gls{csa}}%\gls{ua}
    , \gls{cua} & 0.426 & 0.638 & None \\ 
    \hspace{1em} \makecell[tl]{$\text{SA}^{\textrm{STD}}$,$\text{UA}^{\textrm{STD}}$} & 0.426 & 0.638 & None \\ 
    \hspace{1em} \gls{wsa}, \gls{wua} & 0.085 & 0.117 & Standard normal ability distribution \\

    \bottomrule
\end{tabular}
}
\label{tab:cutoffs:uniform}
\end{table}

To further illustrate the precision of classifications and the impact of the competitive guidelines, we compared estimated classifications with the true underlying categories using the area-based effect-size thresholds of 0.426 (negligible/moderate) and 0.638 (moderate/large).  The underlying \gls{ua} was distributed as 33.35\% negligible, 16.67\% moderate, and 49.98\% large. While $|\Delta_\text{MH}|$ and $|\Delta_\text{LR}|$ achieved approximately 60\% overall agreement using existing cut-off values, $\Delta R^2$ correctly classified only 27.41 \% of cases correctly classified overall (51.05\% negligible, 7.75\% moderate, and 18.19\% large) when using cut-off values of 0.035 (negligible/moderate) and 0.07 (moderate/large). In contrast, applying the newly proposed thresholds of 0.01 (negligible/moderate) and 0.021 (moderate/large) substantially improved alignment to 63.12\% of cases correctly classified (37.88\% negligible, 43.38\% moderate, and 86.54\% large). Notably, 23.94\% of \gls{dif} items were not flagged as statistically significant by the \gls{lr} method, especially for small samples, reducing classification precision. Similar, though slightly smaller, improvements in classification were observed for most other measures, with the exception of LR-P-DIF, which showed only limited gains. The area-based effect-size measures with the proposed cut-off values achieved strong agreement, exceeding 61\% overall (Table~\ref{tab:classification:uniform}). 

\begin{table}[H]
\centering
\caption{Percentage of correctly classified cases using current and newly proposed thresholds for uniform DIF. } 
\label{tab:classification:uniform}
\resizebox{\textwidth}{!}{%
\begin{tabular}{lrrrrrrrr}
  \toprule 
 \multirow{2}{*}{Effect size measure} & \multicolumn{4}{c}{Current classification [\%]} & \multicolumn{4}{c}{Proposed classification [\%]} \\ \cmidrule(lr){2-5} \cmidrule(lr){6-9} 
 & Overall  & Negligible & Moderate & Large & Overall & Negligible & Moderate & Large \\ 
  \midrule
$| \Delta_\text{MH} |$ & 59.31 & 33.49 & 41.33 & 82.52 &  &  &  &  \\ 
$|\hat{\beta}|$ & 41.87 & 16.21 & 4.37 & 71.50 & 51.26 & 33.27 & 33.86 & 69.06 \\ 
$\Delta_\text{LR} $ & 62.08 & 34.45 & 41.63 & 87.34 &  &  &  &  \\ 
$\Delta R^2$ & 27.41 & 51.05 & 7.75 & 18.19 & 63.12 & 37.88 & 43.38 & 86.54 \\ 
LR-P-DIF & 45.50 & 8.45 & 17.39 & 79.59 & 48.38 & 34.02 & 28.10 & 64.73 \\ 
LR-STD-P-DIF total & 52.28 & 13.50 & 18.79 & 89.32 & 61.82 & 37.05 & 35.28 & 87.20 \\ 
LR-STD-P-DIF reference & 52.01 & 13.20 & 17.34 & 89.47 & 61.74 & 36.50 & 34.46 & 87.68 \\ 
LR-STD-P-DIF Cochran & 52.04 & 13.24 & 17.66 & 89.40 & 61.53 & 36.41 & 34.18 & 87.42 \\ 
LR-STD-P-DIF uniform & 56.14 & 17.32 & 37.62 & 88.22 & 63.06 & 38.45 & 51.92 & 83.20 \\ 
UA &  &  &  &  & 63.91 & 41.08 & 52.22 & 83.05 \\ 
WUA &  &  &  &  & 61.58 & 36.66 & 34.20 & 87.33 \\ 
$\text{UA}^{\textrm{STD}}$ &  &  &  &  & 62.09 & 34.51 & 41.54 & 87.35 \\ 
\bottomrule
\end{tabular}
}
\end{table}

%-----------------------
\subsubsection{Non-uniform DIF}

\begin{figure}[!ht]
    \centering
    \caption{Dependencies of effect-size measures on sample size with classification cut-offs indicated by red dashed lines.}
    \begin{subfigure}[b]
    {0.325\textwidth}
        \centering
        \caption{\gls{ua} with cut-offs of 0.426 and 0.638.}
        \includegraphics[width=\textwidth]{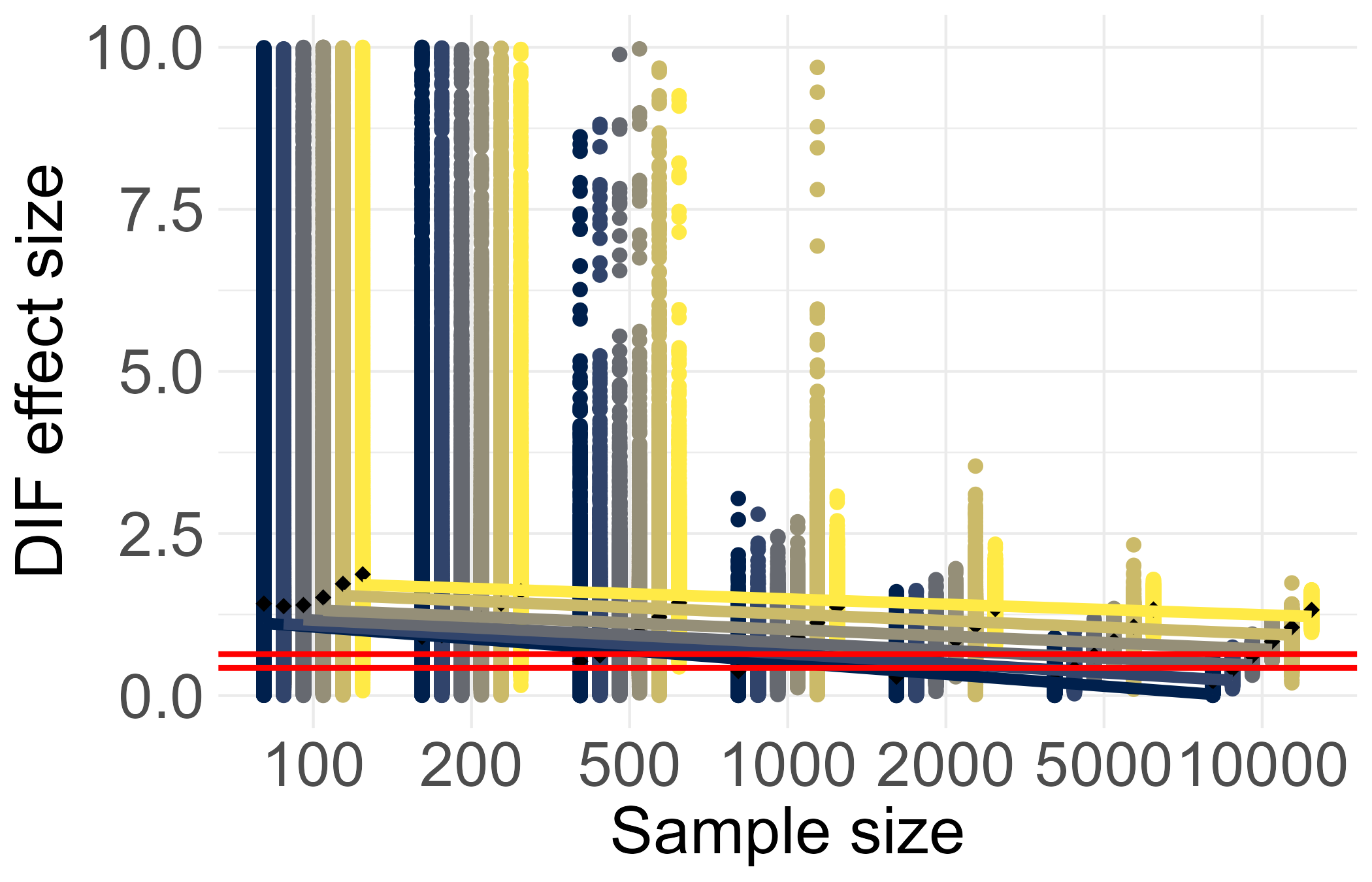}
        \label{fig:nonuniform_ES_samplesize_UA}
    \end{subfigure}
    \hfill
    \begin{subfigure}[b]
    {0.325\textwidth}
        \centering
        \caption{SIBTEST $|\hat{\beta}|$ with cut-offs of 0.059 and 0.088.}
        \includegraphics[width=\textwidth]{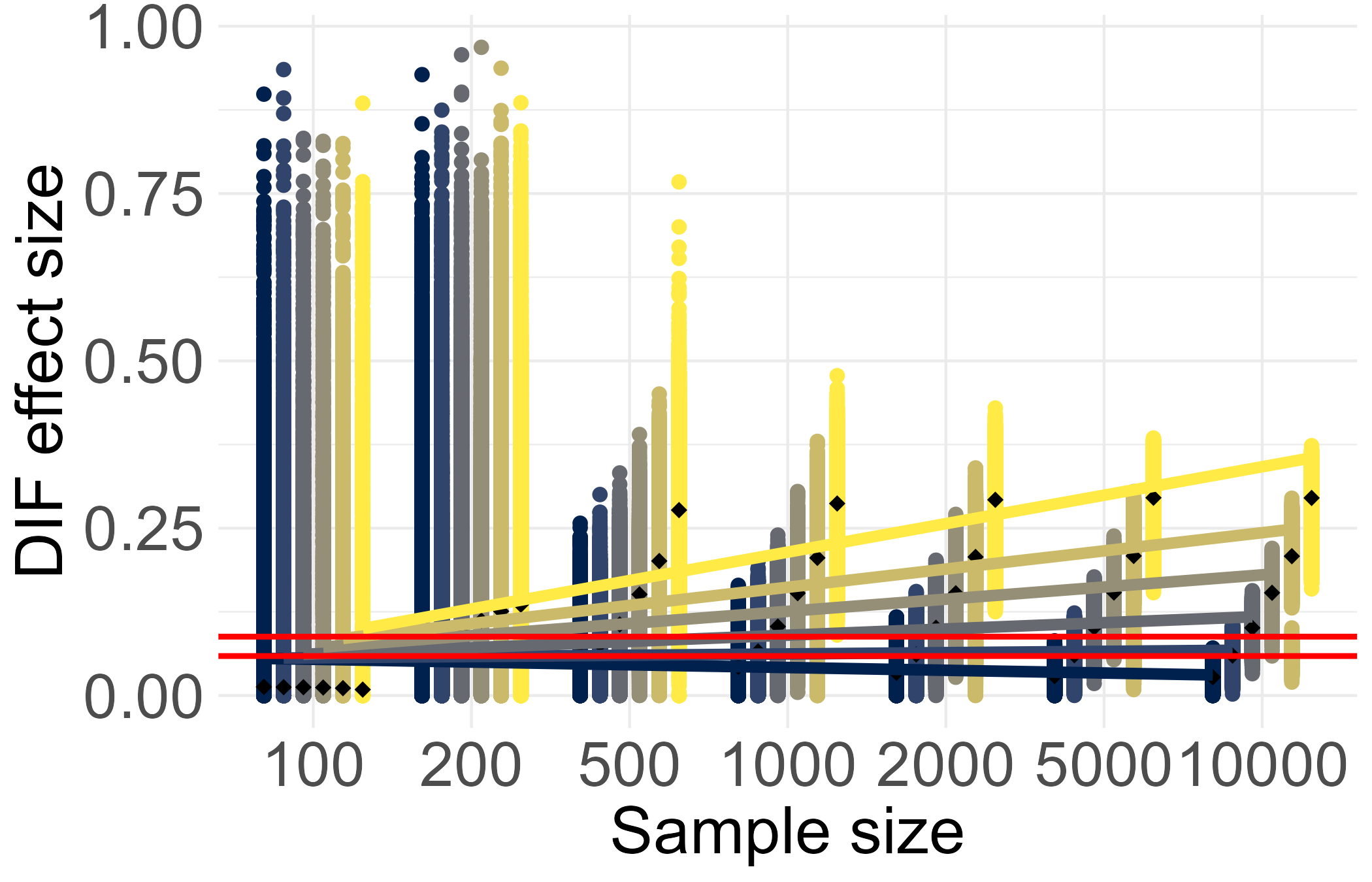}
        \label{fig:nonuniform_ES_samplesize_SIBTEST}
    \end{subfigure}
    \hfill
    \begin{subfigure}[b]{0.325\textwidth}
        \centering
        \caption{$\Delta R^2$ with cut-offs of 0.035 and 0.07.}
        \includegraphics[width=\textwidth]{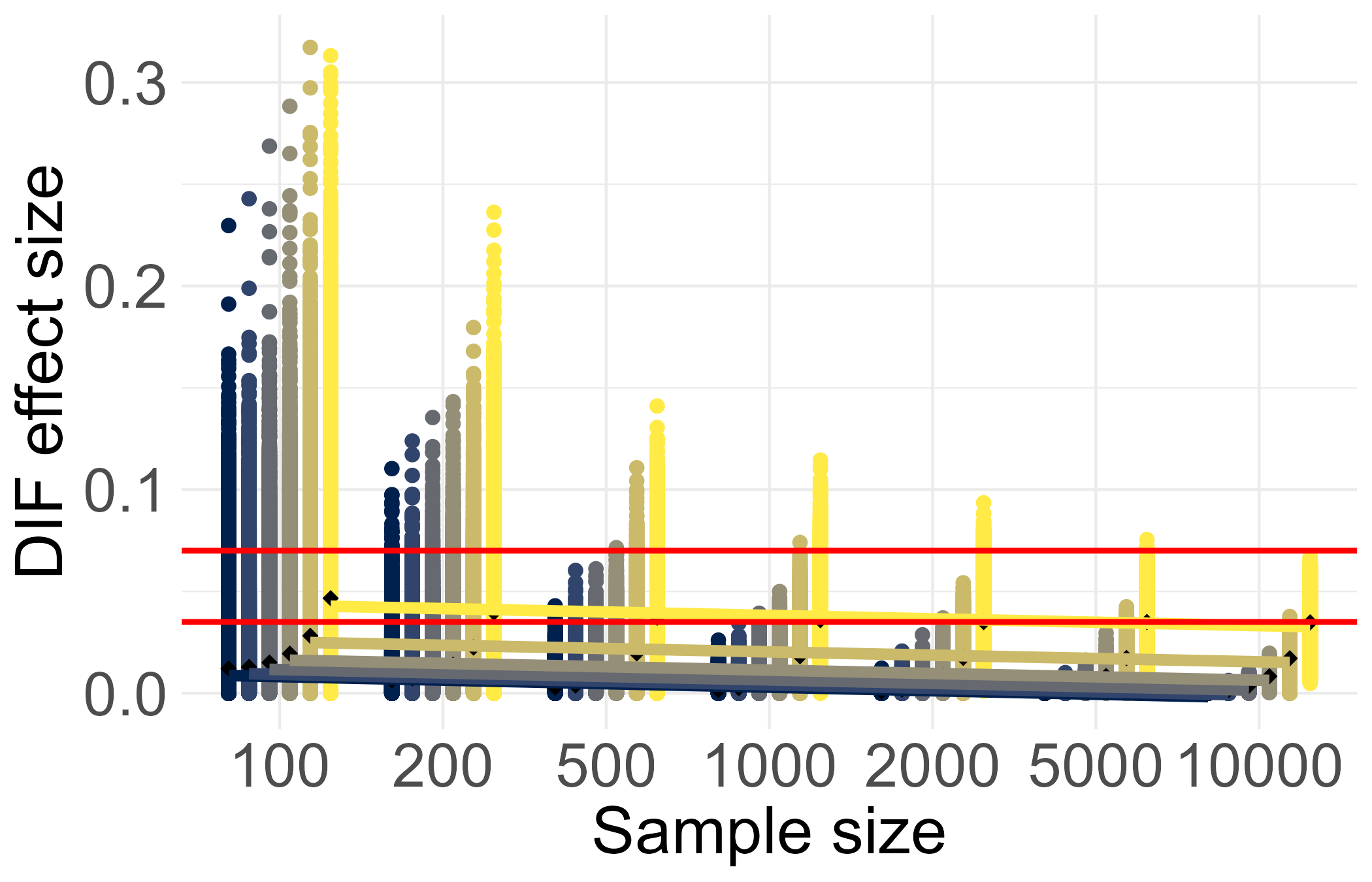}
        \label{fig:nonuniform_ES_samplesize_deltaR2}
    \end{subfigure}
    
    \vspace{-0.6cm} 

    \includegraphics[width=0.75\textwidth]{legend_plot.png}  
    
    \label{fig:nonuniform_ES_samplesize}
\end{figure}

For uniform \gls{dif}, we established classification thresholds of 0.426 and 0.638 for the \gls{ua} measure. To ensure consistency across analyses, these values were also applied in the context of non-uniform \gls{dif} and served as the basis for determining cut-off values for other methods. The same thresholds can be used for the closed version \gls{cua}. Large variance was observed at smaller sample sizes ($N \leq 500$; Figure \ref{fig:nonuniform_ES_samplesize_UA}), a consistent pattern across all effect size measures. Consequently, their use should be restricted to samples with $N \geq 500.$ 

Similarly to \gls{ua}, the crossing \gls{sibtest}'s $|\hat{\beta}|$ exhibited high variability at smaller sample sizes ($N \leq 500$; Figure \ref{fig:nonuniform_ES_samplesize_SIBTEST}), accompanied by relatively weak correlations with the estimated \gls{ua} measure ($\hat{\rho} =$ 0.03--0.5 across sample sizes). To mitigate the influence of these unstable cases, we restricted the analysis to $N \geq 1{,}000$. Within this range, a linear regression model was applied, resulting in thresholds of 0.066 and 0.112 ($\hat{\rho} =$ 0.83--0.95). 

A similar pattern was observed for \gls{wua}, where small sample sizes were associated with high variability and weaker correlations with \gls{ua} ($\hat{\rho} =$ 0.19--0.58). Restricting the analysis to $N \geq 1{,}000,$ a linear regression model produced thresholds of 0.065 and 0.112 ($\hat{\rho} =$ 0.87--0.96). Although some dependence on the underlying ability distribution was observed, its effect on prediction was negligible. 

Consistent with the uniform \gls{dif} setting, the \gls{lr}'s $\Delta R^2$ measure did not exhibit a clear trend with sample size (Figure \ref{fig:nonuniform_ES_samplesize_deltaR2}). A~cubic regression model on \gls{ua} measure was fitted for $N \geq 500$, resulting in cut-off values of 0.002 and 0.008. The preference of a cubic over a linear model was supported by a likelihood ratio test ($p < 0.001$). 

A~summary of the proposed cut-off values, including the corresponding usage restrictions, is provided in Table \ref{tab:cutoffs:nonuniform}.

\begin{table}[!h]
    \centering
    \caption{Newly proposed classification thresholds for effect-size measures used in detecting non-uniform DIF, along with recommended usage restrictions.}
    \begin{tabular}{l cc l}
        \toprule
        \multirow{2}{*}[-0.25em]{\shortstack[l]{DIF method/ \\ \hspace{1em}Effect-size measure}} & 
        \multicolumn{2}{c}{Classification thresholds} &
        \multirow{2}{*}{Usage restrictions} \\
        \cmidrule(lr){2-3}
        & Low & High \\
        \midrule

        \multicolumn{4}{l}{Crossing SIBTEST} \\
        \hspace{1em} $|\hat{\beta}|$ & 0.066 & 0.112 & $N \geq 1{,}000$ \\

        \multicolumn{4}{l}{LR} \\
        \hspace{1em} $\Delta R^2$ & 0.002 & 0.008 & $N \geq 500$ \\ 
        \hspace{1em}\gls{ua}, \gls{cua} & 0.426 & 0.638 & $N \geq 500$ \\
        \hspace{1em}\gls{wua} & 0.065 & 0.112 & $N \geq 1{,}000$ \\

        \bottomrule
    \end{tabular}
    \label{tab:cutoffs:nonuniform}
\end{table}

Similarly to uniform \gls{dif}, we compared the accuracy of effect size classification using both current and newly proposed thresholds. With the existing $\Delta R^2$'s thresholds of 0.035 (negligible/moderate) and 0.07 (moderate/large), only 9\% of cases were correctly classified (21.38\% negligible, 1.54\% moderate, and 3.25\% large). Applying the newly proposed thresholds of 0.002 (negligible/moderate) and 0.008 (moderate/large) yielded a noticeable, though smaller, improvement than in the uniform \gls{dif} setting, with 42.46\% of \gls{dif} items correctly classified (10.93\% negligible, 34.32\% moderate, and 66.2\% large). Notably, the percentage of truly \gls{dif} items which were not flagged as statistically significant by the \gls{lr} method rose to 46.62\%. For the crossing \gls{sibtest} method, this rate was 41.62\%, with the newly proposed thresholds providing only a modest gain of 3.16\% correctly classified cases. The area-based effect-size measures with the proposed cut-off values achieved comparable agreement, exceeding 44\% overall (Table~\ref{tab:classification:nonuniform}). 

\begin{table}[H]
\centering
\caption{Percentage of correctly classified cases using current and newly proposed thresholds for non-uniform DIF. } 
\label{tab:classification:nonuniform}
\resizebox{\textwidth}{!}{%
\begin{tabular}{lrrrrrrrr}
  \toprule 
  \multirow{2}{*}{Effect-size measure} & \multicolumn{4}{c}{Current classification [\%]} & \multicolumn{4}{c}{Proposed classification [\%]} \\ \cmidrule(lr){2-5} \cmidrule(lr){6-9} 
  & Overall  & Negligible & Moderate & Large & Overall & Negligible & Moderate & Large \\ 
  \midrule
$|\hat{\beta}|$ & 42.70 & 17.60 & 13.17 & 69.29 & 45.86 & 21.46 & 31.16 & 67.04 \\ 
$\Delta R^2$ & 9.00 & 21.38 & 1.54 & 3.25 & 42.46 & 10.93 & 34.32 & 66.20 \\ 
UA &  &  &  &  & 44.20 & 9.88 & 22.34 & 74.37 \\ 
WUA &  &  &  &  & 44.96 & 12.71 & 27.67 & 72.21 \\ 
   \bottomrule
\end{tabular}
}
\end{table}

%-------------------------------------------------------------------
% REAL DATA EXAMPLE
%-------------------------------------------------------------------
\section{REAL DATA EXAMPLES}\label{sec:realdata}

The practical utility of the proposed effect size classification is demonstrated using two real-data examples. The first example is drawn from a large-scale study, highlighting the issue of type I error inflation and demonstrating how the use of effect-size measures can help mitigate the detection of trivial differences. The second example comes from a study with a substantially smaller sample size and, consequently, lower statistical power. Together, these examples underscore the need for a~unified classification framework, which enables more consistent and comparable conclusions across different \gls{dif} detection methods.

\subsection{HBSC dataset}
The first example is the \gls{hbsc} dataset, specifically focusing on the multiple health complaints measure items. This dataset originates from the \gls{hbsc} study \cite<see, e.g.,>[]{roberts2009health}, and it includes self-reported data from adolescents aged 11, 13, and 15 on their experiences with a range of symptoms over the past six months. The data have been collected at four-year intervals over the past 40 years, with the most recent cycle conducted in 2022 and used for this analysis. The reported symptoms include headaches, stomach pain, back pain, feelings of low mood, irritability or bad temper, nervousness, difficulty sleeping, and dizziness. Participants reported the frequency of these symptoms using response options ranging from "almost every day" to "rarely or never." 

%-------------------------------------------------------------------
\subsubsection{Data description}

For our analysis, we grouped the responses "almost every day" and "several times a week" into one category (factor level 1), while the remaining responses were combined into another category (factor level 0). \gls{dif} detection was made with respect to the age of respondents, comparing 15-year-olds (reference group) and 11-year-olds (focal group). The aim was to provide granular item-level analysis of group differences in reporting individual symptoms that contribute to overall health. The dataset contains responses from 180{,}553  participants in the year 2022, of which 92{,}198 were 15-year-olds, and 88{,}355 were 11-year-olds. 

%-------------------------------------------------------------------
\subsubsection{Statistical analysis}

\gls{dif} analysis was performed using the \gls{mh} test and the \gls{lr} model \eqref{eq:lr} using likelihood ratio tests to evaluate the presence of uniform and non-uniform \gls{dif} (Table \ref{tab:tests}, second and third rows). The effect size of uniform \gls{dif} was estimated using the following selected measures: $\vert \Delta_{\text{MH}i} \vert$ \eqref{eq:mh:es}, $\Delta R^2$ \eqref{eq:lr:R2diff}, and LR-STD-P-DIF \eqref{eq:lr:LRSTDPDIF} with total and uniform weights, the \gls{ua} measure \eqref{UA} and the $\text{UA}^{\textrm{STD}}$ \eqref{eq:irt:es:UAstd}. The effect size in the case of non-uniform \gls{dif} was estimated using $\Delta R^2$ \eqref{eq:lr:R2diff} and \gls{ua} measure \eqref{UA}. We opted to include well-supported effect sizes based on the simulation results that do not require additional assumptions. The classification of effect sizes was made using both the existing and the newly proposed thresholds (Table~\ref{tab:cutoffs:nonuniform}). In all cases, we used a standardized total score computed from the original ordinal items as the matching criterion to obtain a more precise estimate of the underlying trait, given the low number of items and distributional properties of the total score. For this reason, the \gls{sibtest} with its effect-size measure $\hat{\beta}$ \eqref{eq:sibtest:hatbeta} was not considered. All analyses were conducted at a~5\% significance level. 

%-------------------------------------------------------------------
\subsubsection{Results}

%-----------------------
\paragraph{Uniform DIF}Six items from the \gls{hbsc} dataset (\textit{Backache}, \textit{Dizzy}, \textit{Feel low}, \textit{Nervous}, \textit{Sleep difficulties}, and \textit{Stomachache}) were identified as uniform \gls{dif} items by both the \gls{mh} test and likelihood ratio tests on the \gls{2pl} \gls{lr} model (see Table~\ref{app:tab:example:parameters} for likelihood ratio test results). Among these, three items (\textit{Feel low}, \textit{Irritable}, and \textit{Nervous}) were consistently classified as exhibiting negligible \gls{dif} across all effect-size measures using either current or newly proposed cut-off values. 

For \textit{Backache}, the \gls{dif} magnitude was classified as moderate by the LR-STD-P-DIF measure with uniform weights using existing classification guidelines, while all other effect-size measures indicated negligible \gls{dif}. When applying the newly proposed thresholds, all measures consistently identified this item as negligible, demonstrating improved consistency across methods (Table~\ref{tab:example:merged}). 

\begin{figure}[!ht]
    \centering
    \caption{Estimated ICCs for selected uniform DIF items from the HBSC dataset.}
    \begin{subfigure}[b]{0.495\textwidth}
        \centering
        \caption{Sleep difficulties}
        \includegraphics[width=\textwidth]{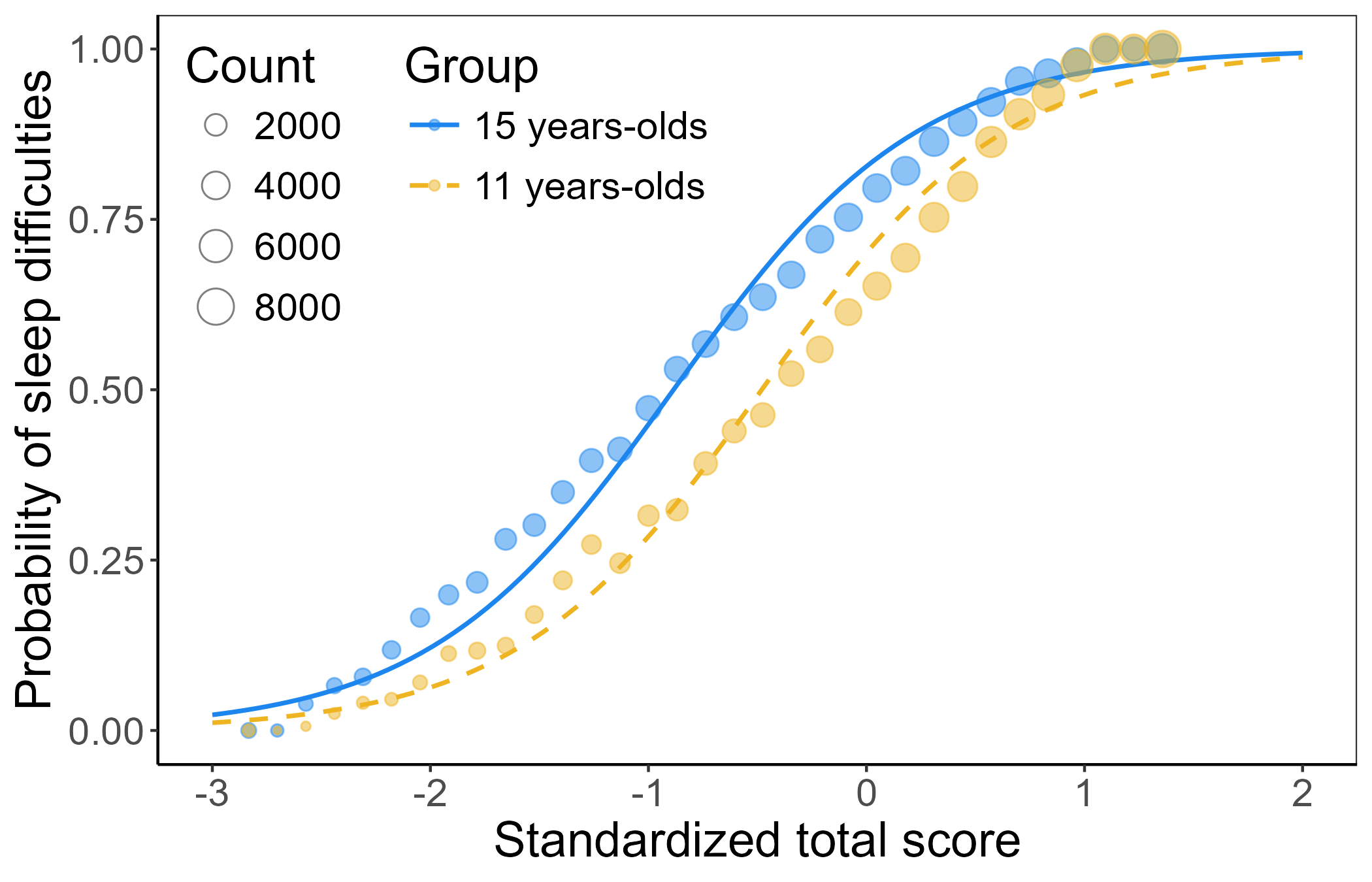}
        \label{fig:example:uniform:sleepdifficulties}
    \end{subfigure}
    \hfill
    \begin{subfigure}[b]{0.495\textwidth}
        \centering
        \caption{Stomachache}
        \includegraphics[width=\textwidth]{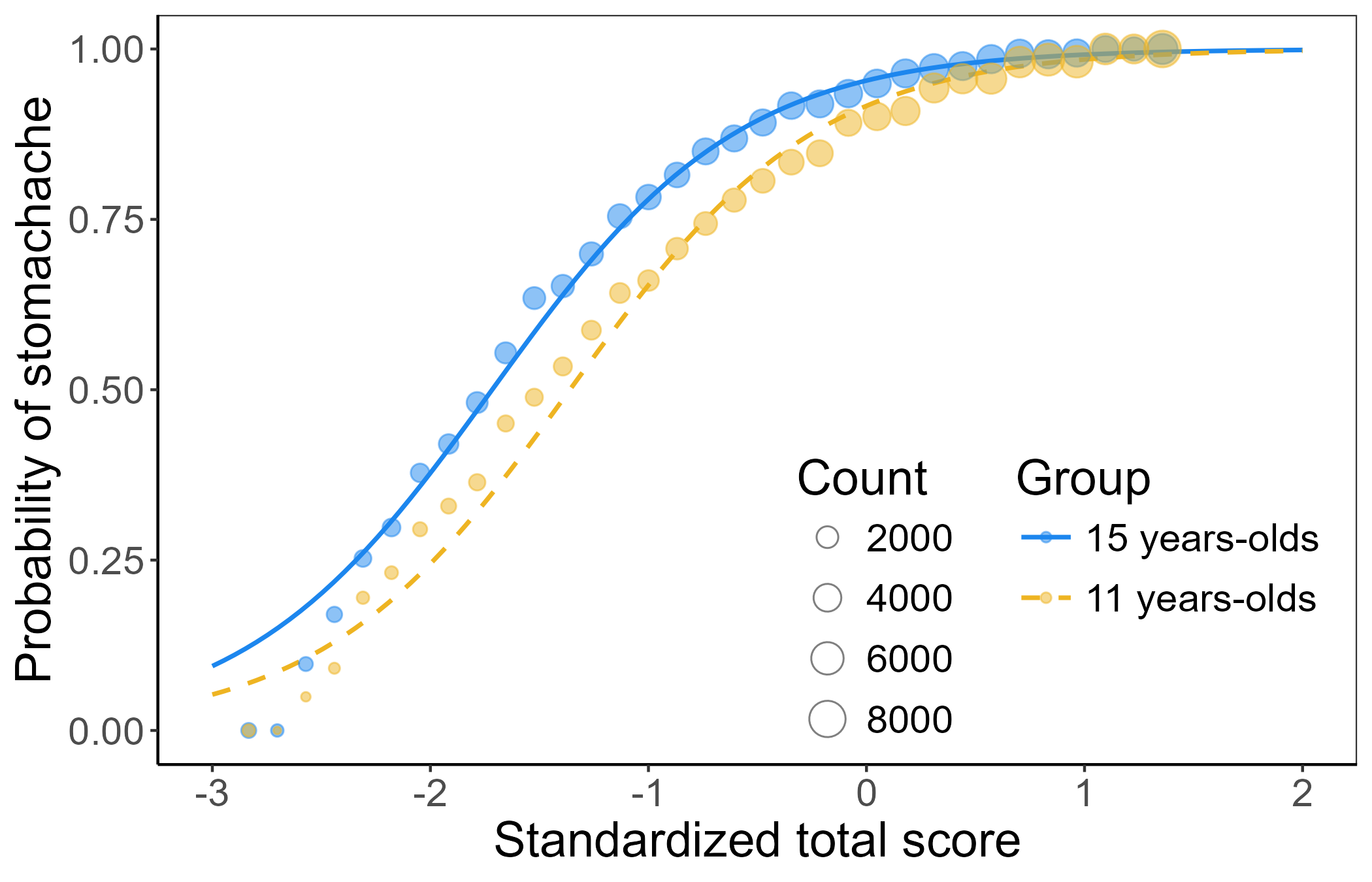}
        \label{fig:example:uniform:stomachache}
    \end{subfigure}
    \label{fig:example:uniform}
    \vspace{-1.5em}
\end{figure}

For \textit{Sleep difficulties}, existing guidelines produced substantial variation in classification across effect-size measures. Specifically, $\Delta R^2$ indicated negligible \gls{dif}, LR-STD-P-DIF with both total and uniform weights classified it as moderate, and $\Delta_{\text{MH}}$ and $\Delta_{\text{LR}}$ as large. In contrast, the newly proposed thresholds brought greater coherence: $\text{UA}^{\textrm{STD}}$ classified the underlying \gls{dif} as large, consistent with the odds-ratio-based measures $\Delta_{\text{MH}}$ and $\Delta_{\text{LR}}$, all of which had values near the moderate/large boundary. $\Delta R^2$ shifted upward from negligible to moderate, while LR-STD-P-DIF with total weights shifted downward from moderate to negligible, although close to the negligible/moderate boundary. Practically, this means that 15-year-old respondents were more likely to report sleeping difficulties than 11-year-olds at the same level of the standardized total score. For example, 11-year-olds with a zero standardized total score had a 70\% probability of reporting sleeping difficulties, while 15-year-olds with the same score had nearly an 83\% probability (Figure~\ref{fig:example:uniform:sleepdifficulties}). 

For \textit{Stomachache}, using the existing cut-off values, the effect-size measures $\Delta_{\text{MH}}$, $\Delta_{\text{LR}}$, and LR-STD-P-DIF with uniform weights classified the underlying \gls{dif} as moderate, while the remaining effect-size measures classified it as negligible. Under the newly proposed thresholds, $\text{UA}^{\textrm{STD}}$ aligned with $\Delta_{\text{MH}}$ and $\Delta_{\text{LR}}$, classifying the effect size as moderate and close to the upper boundary. LR-STD-P-DIF with uniform weights also remained in the moderate range. The remaining measures continued to classify the effect size as negligible, although the $\Delta R^2$ value was near the negligible/moderate boundary. In practical terms, this indicates that, similarly to the \textit{Sleep difficulties} item, 15-year-old respondents were more likely to report stomachache than 11-year-olds at the same standardized total score. For instance, respondents with a standardized total score of $-1$ had a 78\% probability of reporting a stomachache at age 15, compared to 65\% at age 11 (Figure~\ref{fig:example:uniform:stomachache}).

Altogether, \gls{dif} analysis of \gls{hbsc} self-reported symptoms (ages 11 vs. 15) indicated that the items \textit{Sleep difficulties} and \textit{Stomachache} were not age-invariant. 

\begin{table}[H]
\centering
\caption{Classification of selected DIF items in the HBSC dataset. } 
\label{tab:example:merged}
\resizebox{\textwidth}{!}{%
\begin{tabular}{lrrrrrr}
  \toprule 
 & \multicolumn{3}{c}{Uniform DIF} &  \multicolumn{3}{c}{Non-uniform DIF} \\ \cmidrule(lr){2-4} \cmidrule(lr){5-7} 
 \multirow{2}{*}{Effect-size measure} & \multirow{2}{*}{ES value} & \multicolumn{2}{c}{Classification} & \multirow{2}{*}{ES value} & \multicolumn{2}{c}{Classification} \\ \cmidrule(lr){3-4} \cmidrule(lr){6-7} 
 & & \multicolumn{1}{c}{Current} & \multicolumn{1}{c}{Proposed} && \multicolumn{1}{c}{Current} & \multicolumn{1}{c}{Proposed} \\ 
  \midrule
 Backache \\ 
 \midrule
$\Delta_{\text{MH}}$ & 0.661 & Negligible & Negligible &  &  &  \\ 
  $\Delta_{\text{LR}}$ & 0.665 & Negligible & Negligible &  &  &  \\ 
  $\Delta R^2$ & 0.002 & Negligible & Negligible & 0.000 & Negligible & Negligible \\ 
  LR-STD-P-DIF total & 0.036 & Negligible & Negligible &  &  &  \\ 
  LR-STD-P-DIF uniform & 0.054 & Moderate & Negligible &  &  &  \\ 
  UA & 0.188 &  & Negligible & 0.136 &  & Negligible \\ 
  $\text{UA}^{\text{STD}}$ & 0.283 &  & Negligible & &  &  \\ 
   \midrule
 Sleep difficulties \\ 
 \midrule
$\Delta_{\text{MH}}$ & 1.655 & Large & Large &  &  &  \\ 
  $\Delta_{\text{LR}}$ & 1.687 & Large & Large &  &  &  \\ 
  $\Delta R^2$ & 0.015 & Negligible & Moderate & 0.001 & Negligible & Negligible \\ 
  LR-STD-P-DIF total & 0.078 & Moderate & Negligible &  &  &  \\ 
  LR-STD-P-DIF uniform & 0.080 & Moderate & Moderate &  &  &  \\ 
  UA & 0.404 &  & Negligible & 0.462 &  & Moderate \\ 
  $\text{UA}^{\text{STD}}$ & 0.718 &  & Large &  &  &  \\ 
   \midrule
 Stomachache \\ 
 \midrule
$\Delta_{\text{MH}}$ & 1.447 & Moderate & Moderate &  &  &  \\ 
  $\Delta_{\text{LR}}$ & 1.477 & Moderate & Moderate &  &  &  \\ 
  $\Delta R^2$ & 0.010 & Negligible & Negligible & 0.000 & Negligible & Negligible \\ 
  LR-STD-P-DIF total & 0.046 & Negligible & Negligible &  &  &  \\ 
  LR-STD-P-DIF uniform & 0.078 & Moderate & Moderate &  &  &  \\ 
  UA & 0.357 &  & Negligible & 0.311 &  & Negligible \\ 
  $\text{UA}^{\text{STD}}$ & 0.628 &  & Large &  &  &  \\ 
   \bottomrule
\end{tabular}
}
\end{table}

%-----------------------
\paragraph{Non-uniform DIF} The same six items were also flagged as non-uniform \gls{dif} using likelihood ratio tests under the \gls{2pl} \gls{lr} model (see Table~\ref{app:tab:example:parameters}). Closer examination revealed that, under both existing and newly proposed classification guidelines, all effect sizes were classified as negligible. The only exception was the \textit{Sleep difficulties} item, for which the \gls{ua} value of 0.462 is in the lower end of the moderate range (Table \ref{tab:example:merged}).

\subsection{MSATB Dataset}

The second example is the \gls{msatb} dataset, which contains 20 items of the admission test in Biology to medical school in the Czech Republic \cite{drabinova2017detection}, which is available in the \texttt{difNLR} \cite{hladka2020difnlr} and \texttt{ShinyItem\-Analysis} \cite{martinkova2018shinyitemanalysis} \texttt{R} packages. 

\subsection{Data description and statistical analysis}

The dataset consists of dichotomously scored responses from 1{,}407 students. Each item offered four response options and allowed for multiple correct answers. An item was scored as correct only if all correct options and none of the incorrect options were selected. The dataset contains a subset of the original \gls{msatb} items, including \textit{Item49}, which is related to childhood diseases and has been shown to exhibit \gls{dif} by gender in previous studies. 

The \gls{dif} analysis was conducted with respect to the gender of students, comparing males (484) and females (923) using the \gls{mh} test, \gls{sibtest}, and the likelihood-ratio test under the \gls{lr} framework. Similarly to the first example, we opted to include effect sizes that do not require additional assumptions, i.e. $\vert \Delta_{\text{MH}i} \vert$ \eqref{eq:mh:es}, $\vert \hat{\beta} \vert$ \eqref{eq:sibtest:hatbeta}, $\Delta R^2$ \eqref{eq:lr:R2diff}, and LR-STD-P-DIF \eqref{eq:lr:LRSTDPDIF} with total and uniform weights, the \gls{ua} measure \eqref{UA} and the $\text{UA}^{\textrm{STD}}$ \eqref{eq:irt:es:UAstd} in the case of uniform \gls{dif} and $\vert \hat{\beta} \vert$ \eqref{eq:sibtest:hatbeta:nunif} $\Delta R^2$ \eqref{eq:lr:R2diff} and \gls{ua} measure \eqref{UA} in the case of non-uniform \gls{dif}. All analyses were conducted at a~5\% significance level.

\subsubsection{Results}

\paragraph{Uniform DIF} Two items (\textit{Item49} and \textit{Item68}) were identified as uniform \gls{dif} items by both the \gls{mh} test and the likelihood ratio test using the \gls{2pl} \gls{lr} model (see Table \ref{tab:msatb:lr}). The \gls{sibtest} method identified \textit{Item49} as uniform \gls{dif} item, but did not identify \textit{Item68} as significant ($p = 0.076$). 

For \textit{Item49}, the existing guidelines displayed substantial variation in classification across different effect-size measures (Table~\ref{tab:msatb}). Specifically, $\Delta R^2$ indicated negligible \gls{dif}, while LR-STD-P-DIF (with both total and uniform weights), $\Delta_{\text{MH}}$, and $\hat{\beta}$ classified the effect as moderate, and  $\Delta_{\text{LR}}$ as large. In contrast, the newly proposed classification guidelines yielded greater consistency: $\Delta R^2$ and \gls{ua} classified the effect size as moderate in alignment with $\Delta_{\text{MH}}$, $\hat{\beta}$, and LR-STD-P-DIF with uniform weights. Meanwhile,  $\Delta_{\text{LR}}$ and $\text{UA}^{\text{STD}}$ classified the effect size as large, with magnitudes close to the moderate/large cut-off.

For \textit{Item68}, the current classification guidelines resulted in a moderate effect according to LR-STD-P-DIF with total weights, while all other effect-size measures indicated negligible \gls{dif}. In contrast, the newly proposed classification yielded a consistent conclusion: all effect-size measures classified the effect as negligible. 

\paragraph{Non-uniform DIF} The \gls{lr} method did not flag any items as exhibiting potential \gls{dif}. In contrast, the crossing \gls{sibtest} identified statistically significant \gls{dif} for \textit{Item38}, \textit{Item49}, and \textit{Item76}. Under the existing classification guidelines, all three items were classified as having moderate effect sizes. However, when applying the newly proposed cut-off values, the effect sizes for \textit{Item38} and \textit{Item76} were classified as negligible, while \textit{Item49} remained in the moderate category. The newly proposed classification is consistent with the \gls{ua}-based classification (disregarding statistical significance), resulting in greater agreement across methods.

\begin{table}[H]
\centering
\caption{Classification of selected DIF items in the MSATB dataset. } 
\label{tab:msatb}
\resizebox{\textwidth}{!}{%
\begin{tabular}{lrrrrrr}
  \toprule 
 & \multicolumn{3}{c}{Uniform DIF} & \multicolumn{3}{c}{Non-uniform DIF} \\ \cmidrule(lr){2-4} \cmidrule(lr){5-7} 
 \multirow{2}{*}{Effect size measure} & \multirow{2}{*}{ES value} & \multicolumn{2}{c}{Classification} & \multirow{2}{*}{ES value} & \multicolumn{2}{c}{Classification} \\ \cmidrule(lr){3-4} \cmidrule(lr){6-7} 
 & & \multicolumn{1}{c}{Current} & \multicolumn{1}{c}{Proposed} && \multicolumn{1}{c}{Current} & \multicolumn{1}{c}{Proposed} \\ 
  \midrule
 Item38 \\ 
 \midrule
$\Delta_{\text{MH}}$ & 0.471 & Non-significant & Non-significant &  &  &  \\ 
  $|\hat{\beta}|$ & 0.032 & Non-significant & Non-significant & 0.059 & Moderate & Negligible \\ 
  $\Delta_{\text{LR}}$ & 0.475 & Non-significant & Non-significant &  &  &  \\ 
  $\Delta R^2$ & 0.001 & Non-significant & Non-significant & 0.000 & Non-significant & Non-significant \\ 
  LR-STD-P-DIF total & 0.031 & Non-significant & Non-significant &  &  &  \\ 
  LR-STD-P-DIF uniform & 0.024 & Non-significant & Non-significant &  &  &  \\ 
  UA & 0.128 &  & Non-significant & 0.109 &  & Non-significant \\ 
  $\text{UA}^{\text{STD}}$ & 0.201 &  & Non-significant &  &  &  \\ 
   \midrule
 Item49 \\ 
 \midrule
$\Delta_{\text{MH}}$ & 1.435 & Moderate & Moderate &  &  &  \\ 
  $|\hat{\beta}|$ & 0.086 & Moderate & Moderate & 0.086 & Moderate & Moderate \\ 
  $\Delta_{\text{LR}}$ & 1.520 & Large & Large &  &  &  \\ 
  $\Delta R^2$ & 0.016 & Negligible & Moderate & 0.000 & Non-significant & Non-significant \\ 
  LR-STD-P-DIF total & 0.071 & Moderate & Negligible &  &  &  \\ 
  LR-STD-P-DIF uniform & 0.075 & Moderate & Moderate &  &  &  \\ 
  UA & 0.531 &  & Moderate & 0.602 &  & Non-significant \\ 
  $\text{UA}^{\text{STD}}$ & 0.647 &  & Large &  &  &  \\ 
   \midrule
 Item68 \\ 
 \midrule
$\Delta_{\text{MH}}$ & 0.733 & Negligible & Negligible &  &  &  \\ 
  $|\hat{\beta}|$ & 0.049 & Non-significant & Non-significant & 0.049 & Non-significant & Non-significant \\ 
  $\Delta_{\text{LR}}$ & 0.718 & Negligible & Negligible &  &  &  \\ 
  $\Delta R^2$ & 0.004 & Negligible & Negligible & 0.000 & Non-significant & Non-significant \\ 
  LR-STD-P-DIF total & 0.055 & Moderate & Negligible &  &  &  \\ 
  LR-STD-P-DIF uniform & 0.047 & Negligible & Negligible &  &  &  \\ 
  UA & 0.273 &  & Negligible & 0.259 &  & Non-significant \\ 
  $\text{UA}^{\text{STD}}$ & 0.306 &  & Negligible &  &  &  \\ 
   \midrule
 Item76 \\ 
 \midrule
$\Delta_{\text{MH}}$ & 0.464 & Non-significant & Non-significant &  &  &  \\ 
  $|\hat{\beta}|$ & 0.043 & Non-significant & Non-significant & 0.059 & Moderate & Negligible \\ 
  $\Delta_{\text{LR}}$ & 0.470 & Non-significant & Non-significant &  &  &  \\ 
  $\Delta R^2$ & 0.002 & Non-significant & Non-significant & 0.002 & Non-significant & Non-significant \\ 
  LR-STD-P-DIF total & 0.037 & Non-significant & Non-significant &  &  &  \\ 
  LR-STD-P-DIF uniform & 0.029 & Non-significant & Non-significant &  &  &  \\ 
  UA & 0.157 &  & Non-significant & 0.229 &  & Non-significant \\ 
  $\text{UA}^{\text{STD}}$ & 0.200 &  & Non-significant &  &  &  \\ 
   \bottomrule
\end{tabular}
}
\end{table}

%-------------------------------------------------------------------
% DISCUSSION
%-------------------------------------------------------------------
\section{DISCUSSION}\label{sec:discussion}

This work introduced a unified framework for classifying \gls{dif} effect sizes across multiple non-\gls{irt} \gls{dif} detection methods through revised and harmonized cut-off values. We established links among several effect-size measures through both analytical methods and an extensive simulation study, including $\Delta_{\text{MH}}$ for the Mantel-Haenszel test, $\Delta_{\text{LR}}$ for the logistic regression, and area-based measures -- \gls{sa}, and \gls{ua}, as well as the newly proposed standardized versions, $\text{SA}^{\textrm{STD}}$ and $\text{UA}^{\textrm{STD}}$ -- for model-based approaches. 

Notably, we proposed new threshold values of 0.426 and 0.638 for the area-based measures, which distinguish small, moderate, and large levels of \gls{dif}. While previous studies have employed area measures primarily for simulation purposes (e.g., using item areas of 0.4, 0.6, or 0.8, to represent small, medium, or large \gls{dif} \cite<see>{french2007iterative, gomez2009efficacy, drabinova2017detection}, no theory- or simulation-based classification guidelines have been available for area-based \gls{dif} effect-size measures. This study, therefore, fills an important research gap by providing analytically derived classification guidelines for area-based measures, with new thresholds explicitly aligned with the conventional $|\Delta_{\text{MH}}|$'s cut-offs of 1.0 and 1.5.

Additionally, this study empirically evaluated the properties of the effect-size measures. Among all measures considered, those based on area between \glspl{icc} demonstrated the greatest robustness to external factors. Within this group, the standardized area measures $\text{UA}^{\textrm{STD}}$ and $\text{SA}^{\textrm{STD}}$ appear preferable to $\text{UA}$ and $\text{SA}$ for uniform \gls{dif}, as they account for the item discrimination power of items and are directly proportional to effect sizes based on odds ratios. For both types of \gls{dif}, the closed-form variants \gls{csa} and \gls{cua} and weighted variants \gls{wsa} and \gls{wua} offer the advantage of providing finite estimates even under \gls{3pl} and \gls{4pl} models with unequal $c$~and $d$ parameters across groups. These extensions are particularly valuable for applied research, such as investigating whether groups differ in guessing or pretending when responding, and they may also help uncover misconceptions or social barriers \cite{hladka2025newiterative}. 

Furthermore, based on our extensive simulation study, we propose specific usage restrictions. In particular, we advise using the \gls{sibtest}'s $|\hat{\beta}|$ and \gls{ua} only with larger sample sizes ($N \geq 500$ for uniform \gls{dif} and $N \geq 1{,}000$ for non-uniform \gls{dif}), due to its limited power and inconsistent performance in smaller samples. For non-uniform \gls{dif}, other effect-size measures also exhibited high variability at small sample sizes; therefore, larger samples are required for reliable cut-off determination. As a~consequence, we recommend using $\Delta R^2$ only when $N \geq 500$ and \gls{wua} when $N \geq 1{,}000.$ We also advise caution when using effect-size measures that rely on distributional assumptions, such as LR-P-DIF and \gls{wua}, because they may produce unreliable results when the actual ability distribution deviates from the assumed model.

The newly proposed cut-off values and usage recommendations are supported by the scale and comprehensiveness of our simulation design, which included 1{,}008 experimental scenarios with 1{,}000 replications each. This substantially exceeds the scope of many previous \gls{dif} detection studies and enables more precise and reliable conclusions regarding method performance and appropriate usage. For comparison, \citeA{jodoin2001evaluating} considered 36 conditions with 100 replications, while more recently \citeA{hladka2024combining} examined 468 conditions with 1{,}000 replications. The increased coverage of design factors in the present study enables more nuanced, evidence-based guidance on the applicability of individual effect-size measures.

We illustrated the proposed framework using two real-data examples: the \gls{hbsc} study, which measures adolescents' health behaviors, and the \gls{msatb} dataset, containing admission test items in Biology. These examples highlight the importance of effect-size measures, particularly in datasets with very large sample sizes, where even negligible differences may reach statistical significance. The revised cutoffs led to greater consistency in classifications across methods, reducing instances where different effect-size measures would suggest conflicting conclusions for the same item. From a practical perspective, this implies that analysts are less likely to encounter ambiguous evidence when deciding on item flagging or retention, particularly for items with effect sizes near classification thresholds. However, the examples also demonstrate that complete alignment of all effect-size measures is not attainable. For $\text{UA}^{\textrm{STD}}$, we were able to derive thresholds analytically, and, as the real-data example showed, its classifications align closely with those based on $|\Delta_{\text{MH}}|$ and $|\Delta_{\text{LR}}|$. For the remaining effect-size measures, analytical derivation of cut-offs was not feasible; thus, we relied on simulations. Given that the modeled relationships are approximate and contain random variation, minor discrepancies in classification are expected.

The importance of the newly proposed thresholds for area-based measures lies in the broad applicability of these measures across model-based \gls{dif} frameworks. Unlike method-specific effect sizes, area-based measures quantify the magnitude of differences between item response functions and can therefore be applied in a wide range of regression and latent-variable models. This includes regression-based approaches that fit separate models for individual items or a single model for all items jointly, may accommodate fixed or random effects, and incorporate hierarchical and other more complex data structures \cite{martinkova2023computational}. The area-based measures are also applicable when alternative matching variables are used, such as pretest scores in analyses of Differential Item Functioning in Change (DIF-C), also referred to as treatment-related \gls{dif}, which aims to identify item-level heterogeneous treatment effects \cite{martinkova2020academic, gilbert2023modeling, student2026causal}. 

While this study focuses on specific methodological settings, it opens several avenues for future research and generalizations. First, the analysis is limited to traditional and well-established \gls{dif} detection approaches for a single binary grouping variable, specifically the \gls{mh} test, the \gls{sibtest} method, and the model-based approaches with focus on \gls{lr}. While these methods remain among the most commonly used in applied research, extending the proposed unification framework to additional \gls{irt}-based \gls{dif} effect size measures \cite<see, e.g.,>{chalmers2023} and evaluating the proposed area-based thresholds under a wider range of \gls{irt} models, represents a natural next step. 

Second, this work focused on binary items, which are prevalent in assessment practice. Future work could generalize the proposed effect-size measures to polytomous or continuous response formats and establish their correspondence with existing binary-item thresholds. Similarly, generalizations to \gls{dif} involving multi-category grouping variables, continuous covariates, or combinations of multiple covariates would enhance the applicability of the unified framework \cite<see, e.g.,>[]{camilli1999application, kim2007dif, suh2016effect}.

Finally, although the simulations covered a broad range of design factors, they cannot represent all possible real-world scenarios. While our simulation study revealed that the proposed guidelines are robust to moderate deviations from the \gls{2pl} assumptions, the recommended cut-off values are primarily developed for and validated within \gls{2pl}-model settings, and their applicability to more general models, such as \gls{3pl} or \gls{4pl}, is not guaranteed and should be considered with caution. Further validation using empirical data from diverse contexts and item characteristics would strengthen the generalizability of the proposed cut-offs, as also emphasized by \citeA{weese2022reevaluating}.

Despite its scope limitations, this study advances the ongoing effort to standardize the interpretation of \gls{dif} effect sizes by linking and updating classification guidelines across traditional \gls{dif} detection methods for binary items. Through comprehensive simulations, we systematically evaluated the performance and limitations of several commonly used effect-size measures, proposed revised cut-off values to enhance consistency and practical interpretability, and formulated clear recommendations for their use in applied research. The findings highlight the robustness of the area-based effect sizes as practical indicators of \gls{dif}. By addressing methodological inconsistencies and offering a unified framework, this work lays the groundwork for more transparent, reliable, and interpretable \gls{dif} analyses in psychological and educational assessment. 

%--------------------------------------------------
% SUPPLEMENTARY MATERIAL
%--------------------------------------------------

\subsection*{Data availability statement: }
\vspace{-0.5em}
Accompanying \texttt{R} scripts, simulation data, results, and figures are available at \href{https://osf.io/9u5t7/}{https://osf.io/9u5t7/}.

%-----------------------------------------------------------------------
% ACKNOWLEDGEMENT
%-------------------------------------------------------------------

\subsection*{Acknowledgement: }
\vspace{-0.5em}
The study was funded by the Czech Science Foundation project ``Complex analysis of educational measurement data to understand cognitive demands of assessment tasks'' grant number 25-16951S, by the project ``Research of Excellence on Digital Technologies and Wellbeing CZ.02.01.01/00/22\_008/0004583'' which is co-financed by the European Union, by the institutional support RVO 67985807, and by the Charles University, project GA UK No. 117224. The authors thank the editor and the anonymous reviewers for their constructive comments on earlier versions of the manuscript. 

Health Behaviour in School-aged Children (HBSC) is an international study carried out in collaboration with the World Health Organization Regional Office for Europe. The International Coordinator was Jo Inchley (University of Glasgow, United Kingdom) for the 2018 and 2022 survey. The Data Bank Manager was Professor Oddrun Samdal (University of Bergen). We thank Petr Baďura and Jana Furstová for providing helpful consultations on the HBSC dataset.

%-----------------------------------------------------------------------
% REFERENCES
%-------------------------------------------------------------------

\bibliographystyle{apacite}
\bibliography{references}

%-------------------------------------------------------------------

%-------------------------------------------------------------------

%--------------------------------------------------
% APPENDIX
%--------------------------------------------------
    
\clearpage
\appendix
\appendixpage

%--------------------------------------------------

\setcounter{table}{0}
\renewcommand{\thetable}{A\arabic{table}}
\section{Supplementary tables}

\begin{table}[H]
    \centering
    \caption{Item parameters of the 4PL IRT model used to generate non-DIF items. Parameters are the same for both groups. The main simulation uses c=0, d=1.}
    \label{app:tab:nonDIFparameters}
    \begin{tabular}{c rrrr cc rrrr}
        \toprule
        Item & \multicolumn{1}{c}{$a$} & \multicolumn{1}{c}{$b$} & \multicolumn{1}{c}{$c$} & \multicolumn{1}{c}{$d$} & 
        \multicolumn{1}{c}{} & 
        Item & \multicolumn{1}{c}{$a$} & \multicolumn{1}{c}{$b$} & \multicolumn{1}{c}{$c$} & \multicolumn{1}{c}{$d$} \\
        \cmidrule(lr){1-5} \cmidrule(lr){7-11}
        1   & 0.888 & $-0.140$ & 0.072 & 0.822   &   & 19 & 1.140 & 0.175 & 0.082 & 0.832 \\
        2   & 0.954 & $-0.058$ & 0.197 & 0.947   &   & 20 & 0.905 & $-0.118$ & 0.239 & 0.989 \\
        3   & 1.312 & 0.390  & 0.102 & 0.852   &   & 21 & 0.786 & $-0.267$ & 0.222 & 0.972 \\
        4   & 1.014 & 0.018  & 0.221 & 0.971   &   & 22 & 0.956 & $-0.055$ & 0.173 & 0.923 \\
        5   & 1.026 & 0.032  & 0.235 & 0.985   &   & 23 & 0.795 & $-0.257$ & 0.160 & 0.910 \\
        6   & 1.343 & 0.429  & 0.011 & 0.761   &   & 24 & 0.854 & $-0.182$ & 0.249 & 0.999 \\
        7   & 1.092 & 0.115  & 0.132 & 0.882   &   & 25 & 0.875 & $-0.156$ & 0.164 & 0.914 \\
        8   & 0.747 & $-0.316$ & 0.223 & 0.973   &   & 26 & 0.663 & $-0.422$ & 0.177 & 0.927 \\
        9   & 0.863 & $-0.172$ & 0.138 & 0.888   &   & 27 & 1.168 & 0.209  & 0.136 & 0.886 \\
        10  & 0.911 & $-0.111$ & 0.114 & 0.864   &   & 28 & 1.031 & 0.038  & 0.149 & 0.899 \\
        11  & 1.245 & 0.306  & 0.239 & 0.989   &   & 29 & 0.772 & $-0.285$ & 0.072 & 0.822 \\
        12  & 1.072 & 0.090  & 0.113 & 0.863   &   & 30 & 1.251 & 0.314  & 0.037 & 0.787 \\
        13  & 1.080 & 0.100  & 0.169 & 0.919   &   & 31 & 1.085 & 0.107  & 0.241 & 0.991 \\
        14  & 1.022 & 0.028  & 0.143 & 0.893   &   & 32 & 0.941 & $-0.074$ & 0.226 & 0.976 \\
        15  & 0.889 & $-0.139$ & 0.026 & 0.776   &   & 33 & 1.179 & 0.224  & 0.173 & 0.923 \\
        16  & 1.357 & 0.447  & 0.225 & 0.975   &   & 34 & 1.176 & 0.220  & 0.199 & 0.949 \\
        17  & 1.100 & 0.125  & 0.062 & 0.812   &   & 35 & 1.164 & 0.205  & 0.006 & 0.756 \\
        18  & 0.607 & $-0.492$ & 0.011 & 0.761   &   & 36 & 1.138 & 0.172  & 0.119 & 0.869 \\
        \bottomrule
    \end{tabular}
\end{table}

\begin{table}[H]
\centering
\caption{Item parameters of the 2PL IRT model used to generate DIF items. 
} 
\label{app:tab:DIFparameters}
\begin{tabular}{lrrrrrrrrr}
  \toprule 
 \multirow{2}{*}{ES} & \multirow{2}{*}{Item} & \multicolumn{4}{c}{Uniform DIF} & \multicolumn{4}{c}{Non-uniform DIF} \\ 
                               \cmidrule(lr){3-6} \cmidrule(lr){7-10} 
 &  & $a_\text{R}$ & $a_\text{F}$ & $b_\text{R}$ & $b_\text{F}$ & $a_\text{R}$ & $a_\text{F}$ & $b_\text{R}$ & $b_\text{F}$ \\ 
  \midrule 
0.2 & 1 & $1$ & $1$ & $0$ & $0.2$ & $0.5$ & $0.57$ & $0$ & $0$ \\ 
   & 2 & $1$ & $1$ & $0.5$ & $0.7$ & $0.5$ & $0.57$ & $0.5$ & $0.5$ \\ 
   & 3 & $1$ & $1$ & $-0.5$ & $-0.3$ & $0.5$ & $0.57$ & $-0.5$ & $-0.5$ \\ 
   & 4 & $1.5$ & $1.5$ & $0$ & $0.2$ & $0.4$ & $0.44$ & $0$ & $0$ \\ 
   \midrule 
0.4 & 1 & $1$ & $1$ & $0$ & $0.4$ & $0.5$ & $0.66$ & $0$ & $0$ \\ 
   & 2 & $1$ & $1$ & $0.5$ & $0.9$ & $0.5$ & $0.66$ & $0.5$ & $0.5$ \\ 
   & 3 & $1$ & $1$ & $-0.5$ & $-0.1$ & $0.5$ & $0.66$ & $-0.5$ & $-0.5$ \\ 
   & 4 & $1.5$ & $1.5$ & $0$ & $0.4$ & $0.4$ & $0.5$ & $0$ & $0$ \\ 
   \midrule 
0.6 & 1 & $1$ & $1$ & $0$ & $0.6$ & $0.5$ & $0.79$ & $0$ & $0$ \\ 
   & 2 & $1$ & $1$ & $0.5$ & $1.1$ & $0.5$ & $0.79$ & $0.5$ & $0.5$ \\ 
   & 3 & $1$ & $1$ & $-0.5$ & $0.1$ & $0.5$ & $0.79$ & $-0.5$ & $-0.5$ \\ 
   & 4 & $1.5$ & $1.5$ & $0$ & $0.6$ & $0.4$ & $0.57$ & $0$ & $0$ \\ 
   \midrule 
0.8 & 1 & $1$ & $1$ & $0$ & $0.8$ & $0.5$ & $0.98$ & $0$ & $0$ \\ 
   & 2 & $1$ & $1$ & $0.5$ & $1.3$ & $0.5$ & $0.98$ & $0.5$ & $0.5$ \\ 
   & 3 & $1$ & $1$ & $-0.5$ & $0.3$ & $0.5$ & $0.98$ & $-0.5$ & $-0.5$ \\ 
   & 4 & $1.5$ & $1.5$ & $0$ & $0.8$ & $0.4$ & $0.66$ & $0$ & $0$ \\ 
   \midrule 
1 & 1 & $1$ & $1$ & $0$ & $1$ & $0.5$ & $1.29$ & $0$ & $0$ \\ 
   & 2 & $1$ & $1$ & $0.5$ & $1.5$ & $0.5$ & $1.29$ & $0.5$ & $0.5$ \\ 
   & 3 & $1$ & $1$ & $-0.5$ & $0.5$ & $0.5$ & $1.29$ & $-0.5$ & $-0.5$ \\ 
   & 4 & $1.5$ & $1.5$ & $0$ & $1$ & $0.4$ & $0.27$ & $0$ & $0$ \\ 
   \midrule 
1.2 & 1 & $1$ & $1$ & $0$ & $1.2$ & $0.5$ & $1.89$ & $0$ & $0$ \\ 
   & 2 & $1$ & $1$ & $0.5$ & $1.7$ & $0.5$ & $1.89$ & $0.5$ & $0.5$ \\ 
   & 3 & $1$ & $1$ & $-0.5$ & $0.7$ & $0.5$ & $1.89$ & $-0.5$ & $-0.5$ \\ 
   & 4 & $1.5$ & $1.5$ & $0$ & $1.2$ & $0.4$ & $0.97$ & $0$ & $0$ \\ 
   \bottomrule 
\end{tabular}
\end{table}

\begin{table}[H]
\centering
\caption{Likelihood ratio test results and estimated item parameters in the HBSC dataset.} 
\label{app:tab:example:parameters}
\resizebox{\textwidth}{!}{%
\begin{tabular}{lrrrrrrrrrrr}
  \toprule 
 \multirow{2}{*}{Item} & \multicolumn{5}{c}{Uniform DIF} & \multicolumn{6}{c}{Non-uniform DIF} \\ 
 \cmidrule(lr){2-6} \cmidrule(lr){7-12} 
 & $\chi^2$-value & p-value & $a$ & $b$ & $b_{\text{DIF}}$ 
           & $\chi^2$-value & p-value & $a$ & $b$ & $b_{\text{DIF}}$ & $a_{\text{DIF}}$ \\ 
  \midrule
Headache & $0.022$ & $0.881$ & $1.696$ & $-1.177$ & $0.000$ & $1.444$ & $0.230$ & $1.696$ & $-1.178$ & $0.001$ & $0.000$ \\ 
Stomachache & $1341.894$ & $<0.001$ & $1.762$ & $-1.360$ & $-0.357$ & $32.691$ & $<0.001$ & $1.702$ & $-1.387$ & $-0.311$ & $0.117$ \\ 
Backache & $341.611$ & $<0.001$ & $1.504$ & $-1.558$ & $0.188$ & $32.305$ & $<0.001$ & $1.564$ & $-1.521$ & $0.135$ & $-0.100$ \\ 
Feel low & $78.321$ & $<0.001$ & $2.167$ & $-0.825$ & $0.061$ & $12.671$ & $<0.001$ & $2.125$ & $-0.834$ & $0.073$ & $0.076$ \\ 
Irritable & $26.828$ & $<0.001$ & $2.008$ & $-0.546$ & $0.035$ & $1.544$ & $0.214$ & $2.008$ & $-0.546$ & $0.035$ & $0.000$ \\ 
Nervous & $83.436$ & $<0.001$ & $2.035$ & $-0.525$ & $0.061$ & $5.283$ & $0.022$ & $2.060$ & $-0.520$ & $0.055$ & $-0.044$ \\ 
Sleeping difficulties & $2836.715$ & $<0.001$ & $1.775$ & $-0.480$ & $-0.404$ & $260.718$ & $<0.001$ & $1.932$ & $-0.447$ & $-0.462$ & $-0.281$ \\ 
Dizzy & $0.302$ & $0.582$ & $2.053$ & $-1.374$ & $0.000$ & $3.999$ & $0.046$ & $2.081$ & $-1.363$ & $-0.017$ & $-0.046$ \\ 
\bottomrule
\end{tabular}
}
\end{table}

\begin{table}[H]
\centering
\caption{Likelihood ratio test results and estimated item parameters in the MSATB dataset. } 
\resizebox{\textwidth}{!}{%
\label{tab:msatb:lr}
\begin{tabular}{lrrrrrrrrrr}
  \toprule 
 \multirow{2}{*}{Item} & \multicolumn{5}{c}{Uniform DIF} & \multicolumn{5}{c}{Non-uniform DIF} \\ \cmidrule(lr){2-6} \cmidrule(lr){7-11} 
 & $\chi^2$-value & p-value & $a$ & $b$ & $b_{\text{DIF}}$  & $\chi^2$-value & p-value & $a$ & $b$ & $b_{\text{DIF}}$ \\ 
  \midrule
Item49 & $14.621$ & $<0.001$ & $1.218$ & $-1.510$ & $-0.531$ & $0.139$ & $0.709$ & $1.218$ & $-1.510$ & $-0.531$ \\ 
  Item27 & $1.078$ & $0.299$ & $1.262$ & $1.177$ & $0.000$ & $0.135$ & $0.714$ & $1.265$ & $1.255$ & $-0.122$ \\ 
  Item41 & $0.219$ & $0.640$ & $1.689$ & $0.419$ & $0.000$ & $0.418$ & $0.518$ & $1.690$ & $0.445$ & $-0.040$ \\ 
  Item7 & $1.817$ & $0.178$ & $1.259$ & $0.333$ & $0.000$ & $0.085$ & $0.770$ & $1.259$ & $0.242$ & $0.140$ \\ 
  Item38 & $1.964$ & $0.161$ & $1.574$ & $-0.778$ & $0.000$ & $0.183$ & $0.669$ & $1.580$ & $-0.693$ & $-0.128$ \\ 
  Item28 & $0.691$ & $0.406$ & $1.042$ & $1.722$ & $0.000$ & $0.005$ & $0.945$ & $1.043$ & $1.802$ & $-0.125$ \\ 
  Item9 & $3.117$ & $0.077$ & $1.235$ & $-0.920$ & $0.000$ & $0.291$ & $0.590$ & $1.242$ & $-0.789$ & $-0.196$ \\ 
  Item47 & $1.781$ & $0.182$ & $1.290$ & $-2.409$ & $0.000$ & $2.542$ & $0.111$ & $1.284$ & $-2.583$ & $0.241$ \\ 
  Item75 & $0.058$ & $0.810$ & $1.177$ & $-0.227$ & $0.000$ & $0.007$ & $0.932$ & $1.178$ & $-0.209$ & $-0.026$ \\ 
  Item17 & $2.302$ & $0.129$ & $1.186$ & $1.034$ & $0.000$ & $0.493$ & $0.482$ & $1.190$ & $1.150$ & $-0.181$ \\ 
  Item76 & $2.358$ & $0.125$ & $1.278$ & $-0.245$ & $0.000$ & $3.025$ & $0.082$ & $1.277$ & $-0.348$ & $0.157$ \\ 
  Item10 & $0.005$ & $0.944$ & $1.503$ & $-1.376$ & $0.000$ & $0.310$ & $0.578$ & $1.502$ & $-1.381$ & $0.008$ \\ 
  Item64 & $0.052$ & $0.820$ & $0.745$ & $0.095$ & $0.000$ & $2.177$ & $0.140$ & $0.746$ & $0.119$ & $-0.036$ \\ 
  Item45 & $2.182$ & $0.140$ & $1.278$ & $-0.739$ & $0.000$ & $0.384$ & $0.536$ & $1.278$ & $-0.844$ & $0.158$ \\ 
  Item24 & $1.809$ & $0.179$ & $1.293$ & $0.557$ & $0.000$ & $0.001$ & $0.972$ & $1.293$ & $0.467$ & $0.139$ \\ 
  Item1 & $1.279$ & $0.258$ & $0.785$ & $-0.999$ & $0.000$ & $0.431$ & $0.511$ & $0.784$ & $-1.121$ & $0.183$ \\ 
  Item68 & $5.272$ & $0.022$ & $1.118$ & $-0.939$ & $0.273$ & $0.034$ & $0.853$ & $1.118$ & $-0.939$ & $0.273$ \\ 
  Item61 & $1.751$ & $0.186$ & $1.286$ & $-1.151$ & $0.000$ & $1.868$ & $0.172$ & $1.292$ & $-1.051$ & $-0.149$ \\ 
  Item25 & $1.867$ & $0.172$ & $0.957$ & $-0.527$ & $0.000$ & $0.065$ & $0.799$ & $0.956$ & $-0.646$ & $0.180$ \\ 
  Item2 & $0.472$ & $0.492$ & $1.012$ & $1.190$ & $0.000$ & $0.462$ & $0.497$ & $1.014$ & $1.250$ & $-0.094$ \\ 
   \bottomrule
\end{tabular}
}
\end{table}

{

\begingroup

\section{Sensitivity analysis}
\label{app:sensitivity}

To evaluate the robustness of the proposed classification guidelines, we conducted a sensitivity analysis under a combined modification of the simulation design that simultaneously targets two key sources of potential variability.

First, non-\gls{dif} items were generated using the \gls{4pl} model instead of the \gls{2pl} model used in the main analysis. This modification introduces additional flexibility by adding parameters for guessing and the upper asymptote, thereby relaxing the structural assumptions of the primary design.

Second, the focal-group ability distribution was assumed to be normal, $\mathcal{N}(\mu, 1)$, with $\mu = 0$ and $\mu = -0.5$. While the main simulation study considers symmetric ability differences ($\mu = -0.5, 0, 0.5$), the present analysis focuses on a subset of conditions to examine whether the results depend on the direction and magnitude of the ability shift, and to assess the combined impact of the alternative measurement model and distributional assumptions within a single setting.

The other design factors were chosen analogously to the main simulation study. In total, the simulation study includes 7 (sample sizes) $\times$ 2 (test lengths) $\times$ 2 (proportions of \gls{dif} items) $\times$ 2 (latent ability distributions for the focal group) $\times$ 6 (underlying \gls{dif} effect size values) $\times$ 2 (types of \gls{dif}) $=$ 672 different settings. Each simulation setting was replicated 1{,}000 times.

By jointly varying the measurement model and the ability distribution, this sensitivity analysis provides a stringent stress test of the proposed thresholds. In particular, it allows us to assess whether conclusions remain stable under conditions where both the shape of the item characteristic curve and the distribution of the latent trait deviate from the baseline assumptions.

\subsection{Uniform DIF}

\begin{table}[H]
\centering
\caption{Sensitivity analysis for uniform DIF: Comparison of proposed thresholds under the primary 2PL design and the combined alternative design (4PL with $\mu \neq 0.5$).}
\label{tab:sensitivity_uniform}
\resizebox{\textwidth}{!}{%
\begin{tabular}{lccccl}
\toprule
\textbf{Effect size} & \multicolumn{2}{c}{\textbf{Thresholds}} & & \textbf{Change} & \textbf{Conclusion} \\
\cmidrule(lr){2-3}
 & \gls{2pl} design & \gls{4pl} + $\mu \neq 0.5$ && & \\
\midrule

$|\Delta_{\text{MH}}|$ 
& 1, 1.5 
& 1, 1.5 
&& none 
& Cut-offs not derived \\

$|\hat{\beta}|$ 
& 0.085, 0.122 
& 0.082, 0.118 
&& negligible decrease
& Robust; same usage recommendation \\

$\Delta_{\text{LR}}$ 
& 1, 1.5 
& 1, 1.5 
&& none 
& Robust \\

$\Delta R^2$ 
& 0.010, 0.021 
& 0.011, 0.022 
&& slight increase 
& Robust \\

LR-P-DIF 
& 0.092, 0.122 
& 0.079, 0.093 
&& moderate decrease 
& Sensitive to distribution \\

LR-STD-P-DIF 
& 0.084, 0.116 
& 0.084, 0.116 
&& none 
& Robust \\

\gls{sa}, \gls{ua} 
& 0.426, 0.638 
& 0.426, 0.638 
&& none 
& Cut-offs derived algebraically \\

\gls{sa}$^{\text{STD}}$, \gls{ua}$^{\text{STD}}$ 
& 0.426, 0.638 
& 0.426, 0.638 
&& none 
& Cut-offs derived algebraically \\

\gls{wua} 
& 0.085, 0.117 
& 0.084, 0.116 
&& negligible decrease 
& Robust \\
\bottomrule
\end{tabular}
}
\end{table}

\subsection{Non-uniform DIF}

\begin{table}[H]
\centering
\caption{Sensitivity analysis for non-uniform DIF: Comparison of proposed thresholds under the primary 2PL design and the combined alternative design (4PL with $\mu \neq 0.5$).}
\label{tab:sensitivity_nonuniform}
\resizebox{\textwidth}{!}{%
\begin{tabular}{lccccl}
\toprule
\textbf{Effect size} & \multicolumn{2}{c}{\textbf{Thresholds}} & & \textbf{Change} & \textbf{Conclusion} \\
\cmidrule(lr){2-3}
 & \gls{2pl} design & \gls{4pl} + $\mu \neq 0.5$ && & \\
\midrule

$|\hat{\beta}|$ (crossing) 
& 0.066, 0.112 
& 0.058, 0.085 
&& decrease 
& Sensitive to the underlying model
\\

$\Delta R^2$ 
& 0.002, 0.008 
& 0.008, 0.015 
&& increase
& Sensitive to the underlying model \\

\gls{ua}, \gls{cua} 
& 0.426, 0.638 
& 0.426, 0.638 
&& none 
& Cut-offs taken from uniform \gls{dif} \\

\gls{wua} 
& 0.065, 0.112 
& 0.058, 0.088 
&& decrease
& Sensitive to ability distribution and underlying model \\

\bottomrule
\end{tabular}
}
\end{table}

\subsection{Interpretation}

The results show that the proposed classification framework is largely robust to combined changes in the model and ability distributions. For uniform \gls{dif}, the thresholds remain essentially unchanged across all measures. In particular, only LR-P-DIF shows a moderate decrease (14\% and 24\%), which can be attributed to its sensitivity to the latent trait distribution. For non-uniform \gls{dif}, the cut-offs changed more noticeably, suggesting sensitivity to the underlying data-generating model. 

Overall, for uniform \gls{dif}, the cut-offs remain stable, indicating that the proposed thresholds are not driven by a specific modeling assumption and generalize well beyond the baseline design. For non-uniform \gls{dif}, caution should be exhibited when generalizing the results for other situations, e.g., for \gls{3pl} or \gls{4pl} models.
\endgroup

\end{document}